\documentclass[12pt]{article}
\input{epsf.sty}

\def\mydate{21 Feb 2000}
\def\ignore#1{{}}

\newcommand{\upri}{u^{\prime}}
\newcommand{\wpri}{w^{\prime}}

\newcommand{\beeq}{\begin{equation}}
\newcommand{\eneq}{\end{equation}}
\newcommand{\beqn}{\begin{eqnarray}}
\newcommand{\eeqn}{\end{eqnarray}}

\def\mybig{\displaystyle \strut }

\def\la{\raise.16ex\hbox{$\langle$}\lower.16ex\hbox{}  }
\def\ra{\, \raise.16ex\hbox{$\rangle$}\lower.16ex\hbox{} }
\def\go{\rightarrow}

\def\onehalf{ \hbox{${1\over 2}$} }

\def\psibar{ \psi \kern-.65em\raise.6em\hbox{$-$} \lower.6em\hbox{} }
\def\psibaralpha{ \psi^{(\alpha)} \kern-1.9em\raise.6em\hbox{$-$}
\kern+1.2em\hbox{}}
\def\psibara{ \psi^{(a)} \kern-1.9em\raise.6em\hbox{$-$}\kern+1.2em\hbox{}}

\def\vphi{\varphi}
\def\ep{\epsilon}

\def\myfrac#1#2{{\mybig #1\over \mybig #2}}

\begin{document}

\thispagestyle{empty}

\baselineskip=12pt

{\small \noindent \mydate ~(corrected.) \hfill UMN-TH-1838/00}

{\small  \hfill BNL-NT-00/2}

{\small  \hfill NUC-MINN-00/3-T}

{\small \hfill \bf hep-th/0002098}

\baselineskip=40pt plus 1pt minus 1pt

\vskip 3cm

\begin{center}

{\Large\bf Monopoles, Dyons and Black Holes in the Four-Dimensional
Einstein-Yang-Mills Theory}\\

\ignore{
{\Large\bf Monopoles and Dyons}\\
{\Large\bf  in the Four-Dimensional Einstein-Yang-Mills
Theory}\\
}

\vspace{1.0cm}
\baselineskip=20pt plus 1pt minus 1pt

{\large   Jeff Bjoraker\footnote{Current address:
{\it Brookhaven National Laboratory, Building 510A}  
{\it  Upton, NY 11973, U.S.A.}}
and   Yutaka Hosotani}\\
\vspace{.1cm}
{\it School of Physics and Astronomy, University of Minnesota}\\  
{\it  Minneapolis, MN 55455, U.S.A.}\\ 
\end{center}

\vskip 2.cm
\baselineskip=20pt plus 1pt minus 1pt

\begin{abstract}
A continuum of  monopole, dyon and black hole  solutions exist in
the Einstein-Yang-Mills theory in asymptotically anti-de Sitter space.  
Their structure is studied in detail.  The solutions are classified by
non-Abelian electric and magnetic charges and the ADM mass.   The
stability of the solutions which have no node in non-Abelian magnetic 
fields is established.  There exist critical spacetime solutions
which terminate at a finite radius, and  have
universal  behavior.   The moduli space of the 
solutions exhibits a fractal structure as the cosmological constant
approaches zero.   
\end{abstract}

\newpage


\newpage
\section{Introduction}

For a long time, it was believed that no regular particle-like stable solutions 
(solitons) with
finite mass can exist in self gravitating systems unless the
stability is guaranteed topologically.  The Einstein theory
in vacuum and the Einstein-Maxwell system do not admit solitons. 
It came as quite a surprise when Bartnik and
McKinnon (BK) found globally regular solutions to the $SU(2)$ Einstein
Yang-Mills (EYM) theory without scalar fields\cite{BARTNIK}. It was unexpected
to find that self gravitating Yang-Mills systems produced solitons.
Unfortunately, the BK solutions were shown
to be unstable against linear perturbations \cite{ZHOU}. 
Later, other fields such as Higgs scalar fields and dilaton fields
 were included in the EYM action,  but with the exception of the
Skyrmions, all turned out to be unstable (see \cite{VOLKOV2} for a
review).   

Interest in the BK solutions was renewed with the discovery of black
hole solutions to the EYM equations \cite{BIZON,KUNZLE}.  These
non-Abelian black holes aparently violate the no-hair conjecture
\cite{HEUSLER}.  But these   non-Abelian black hole solutions are also
unstable,  and again other fields were added in the hope of achieving
stability without success (see Ref.'s \cite{VOLKOV2,BREITENLOHNER}) for a
review).

We stress that it is a surprise that there are static solutions to the
Einstein Yang-Mills equations at all.
There are no static solutions to the Yang-Mills equations in 4
dimensional flat space. We can see this with a simple argument given
by Deser \cite{DESER}.
The conservation of the canonical energy momentum tensor,
$\partial_{\nu}T^{\mu\nu} =0$, implies that for a static field
configuration $\partial_{j} {T_i}^{j} =0$. The total
divergence of the quantity $x^i{T_i}^{j}$ must vanish to maintain
finite energy and regularity, 
$\int d^{d-1}x \partial_j(x^i {T_i}^{j}) = 0$.
But $\partial_j(x^i {T_i}^{j}) = {T_j}^j + x^i \partial_{j} {T_i}^{j}
={T_j}^j $ so that
\begin{equation}
\int  d^{d-1}x ~ T^i_i = \int
d^{d-1}x\left[\frac{1}{2}(5-d)F^2_{ij}+(d-3)F^2_{0i}\right] = 0.
\end{equation} 
Since the integrand above is positive definite for $d=4$, $F_{ij}$ and
$F_{0i}$ must vanish. Thus there are no regular static solutions.

The argument above cannot be extended to curved spacetime.  The 
conservation law ${T^{\mu\nu}}_{;\nu}=0$ leads to
\beeq
\int d^{d-1} x \, \sqrt{-g} ~{T_j}^j = 
- \int d^{d-1}x \,\sqrt{-g} ~ x^k   ~\Gamma_{k\mu\nu} T^{\mu\nu}
\not= 0 ~.
\eneq
The failure of Deser's simple
argument in curved space implies the possibility of having static
solutions in curved space. Gravity supplies the attractive force needed to
balance the repulsive force of Yang-Mills gauge interactions. 
Indeed, any solution to $SU(2)$ EYM equations in asymptotically Minkowski space which is regular
asymptotically is also regular for all $r>0$ \cite{SMOLLER}.

The particle-like and black hole solutions were later studied in a
cosmological context. The behavior of static solutions to the Einstein
Yang-Mills equations depends considerably  on the sign of the
cosmological constant. The solutions can be separated into two families;
$\Lambda \ge 0$ and $\Lambda < 0$. The solutions where $\Lambda = 0$ are
the BK solutions. Their asymptotically de Sitter analogs
($\Lambda > 0$) were discovered independently by Volkov et. al. and
Torii et. al. \cite{VOLKOV}. The BK solutions and the cosmological
extensions  all share similar behavior, and are  unstable
\cite{GREENE,BRODBECK}. (See Ref. \cite{VOLKOV2} for a review).  
Recently,  asymptotically anti-de
Sitter black hole solutions \cite{WINSTANLEY} and soliton solutions
\cite{BJORAKER,Bjoraker2} were found which are strikingly different from
the BK type solutions. In particular, the asymptotically anti-de Sitter
AdS EYM equations have solutions where the field strengths  
are non-zero everywhere. These solutions were also
shown to be stable against spherically symmetric linear
perturbations. These solutions are the only EYM solutions solutions
that are stable. This discovery would be very important to cosmology if the
universe was ever in a phase where the cosmological constant is negative.

Another new feature of the EYM theory in AdS is the existence of dyon
solutions. If $\Lambda \ge 0$ the electric part of the gauge fields
is forbidden \cite{GALTSOV} if the ADM mass is to remain
finite. Scalar fields must be added to the theory in order for the
boundary conditions at infinity to permit the electric 
fields and maintain a finite ADM mass \cite{BRIHAYE,LUGO}.  

Recently a tremendous amount of interest has evolved in field theories in
AdS space. There is the AdS/CFT correspondence \cite{MALDACENA}. 
Conformal field theories  in $d$ dimensions ($R_d$) are described in
terms of supergravity or string theory on the product space of
AdS$_{d+1}$ and a compact manifold.  There are intimate relations between
data on the boundary
$R_d$ of AdS$_{d+1}$   and data in the bulk AdS$_{d+1}$.  In the present
paper we are examining the Einstein-Yang-Mills theory in asymptotically
AdS space.  The  boundary in space must be playing 
a crucial role for the existence of stable monopole and dyon solutions,
more detailed analysis of which is, however, left for future
investigation.  We also note that in the three-dimensional 
AdS space there exist nontrivial black holes \cite{BTZ} and 
monopole/instanton solutions \cite{Tekin}.

When the value of the cosmological constant $\Lambda$ is varied, the space
of monopole and dyon solutions, the moduli space, also changes.  With a
finite negative $\Lambda$, solutions exist in continuum. They are
classified in a  finite number of families, or branches.  With a vanishing
or positive
$\Lambda$ solution exists only in a discrete set, but there are infinitely
many.  One natural question emerging is how these finite number of
branches of solutions in continuum become infinitely many discrete
points as $\Lambda<0$ approaches 0.  There is a surprising hidden
feature in this limit.  We shall find a fractal structure in the 
moduli space, which seems to explain the transition. 

In the next section the general formalism is given and the equations
of motion are derived with a spherically symmetric ansatz.  Conserved charges in the 
Yang-Mills theory is defined in section 3.  Some general no-go
theorems are derived from sum rules in Section 4.  New soliton
solution in asymptotically anti-de Sitter space are explained in
section 5.   The critical spacetime which have universality near
the edge of the space is also examined.   Black hole solutions
which have both magnetic and electric non-Abelian charges are
presented in Section 6.  The dependence of the moduli space on
the cosmological constant $\Lambda$ is investigated in Section 7 where
the fractal structure is revealed when $\Lambda$ approaches zero from the 
negative side.  The detailed analysis of the stability of the monopole
solutions is presented in Section 8.  The subtle boundary condition
in the problem requires elaboration of the previous argument
presented in the $\Lambda=0$ and $\lambda>0$ cases.

\section{General Formalism}

In non-Abelian gauge theory, the field equations have solutions which
exhibit a magnetic charge.   In the 't Hooft-Polyakov monopole 
solution
\begin{equation}
\begin{array}{lcr}
\Phi_a = \myfrac{x_a}{er^2} H(er), &A_a^0 = 0, & A^i_a =
-\epsilon_{aij} \myfrac{x_j}{er^2}(1-K(er)).
\end{array}
\label{TP_ANSATZ}
\end{equation}
where $\Phi_a$ is a triplet Higgs scalar field.
Its stability is guaranteed by the topology of the triplet
Higgs scalar field.\cite{THOOFT} The $U(1)$ magnetic charge takes
a quantized value, $4\pi/e$.
Dyon solutions were obtained  \cite{JULIA} starting
with the above ansatz (\ref{TP_ANSATZ}) but with a non-zero value for
$A_a^0$, (i.e. $A_a^0 = (x_a/er^2)J(er)$).

In this paper we look for monopole and dyon solutions in the
Einstein-Yang-Mills theory without scalar fields;  
\beeq
S=\int d^4x\sqrt{-g}  
\left[ \frac{1}{16\pi G}(R-2\Lambda)
- {1\over 4} F^{a\mu\nu} {F^a}_{\mu\nu} \right].
\label{action1}
\eneq
The Einstein and Yang-Mills equations are given by
\beqn
&R^{\mu\nu} - {1\over 2}g^{\mu\nu} (R - 2 \Lambda) = 8\pi G ~
T^{\mu\nu}&\cr
&{F^{\mu\nu}}_{;\mu} + e [A_\mu, F^{\mu\nu} ] = 0&
\label{EYM-eq}
\eeqn
We suspect that the gravity provides attractive force to balance the 
equation.  

We look for spherically symmetric solutions.  The metric takes the form
\beeq
ds^2 = -\myfrac{H}{p^2} \, dt^2 + \myfrac{dr^2}{H} 
 + r^2 (d\theta^2 + \sin^2 \theta \, d\phi^2)
\label{metric1}
\eneq
whereas Yang-Mills fields are given \cite{WITTEN,MANTON}, in the regular
gauge, by
\beqn
A^{(0)} &=& {\tau^j\over 2e} \Bigg\{ 
{A_0}  {x_j\over r}  dt + {A_1}  {x_j  x_k\over r^2} dx_k 
\cr
\noalign{\kern 10pt}
&&\hskip .7cm+{\phi_1\over r} 
 \bigg(\delta_{jk} - {x_j x_k\over r^2} \bigg)  dx_k  
- \ep_{jkl} {1-\phi_2 \over r^2} { x_k} dx_l \Bigg\} 
\label{YM-ansatz1}
\eeqn
Here the Cartesian coordinate $x^k$'s are related to the polar
coordinates $(r,\theta,\phi)$ as in the flat space.
$H$, $p$, $A_0$, $A_1$, $\phi_1$ and $\phi_2$ are  functions
of $r$   for monopole or dyon solutions.  In the discussion of the
stability of the solutions they depend on both $t$ and $r$.
The regularity of solutions at the origin demands that $H$, $p$ are
finite, whereas $A_0$, $A_1$, and $\phi_1 \rightarrow 0$ and
$\phi_2 \rightarrow 1$ at $r=0$.

\subsection{Simplification of the static gauge field ansatz} 

Let $A =A_{\mu}dx^\mu = \frac{1}{2}\tau^a A_{\mu}^a dx^\mu$, where
$\tau^a$ are the usual Pauli matrices.  In terms of the basis in spherical
coordinates 
$(\tau_r, \tau_\theta, \tau_\phi)=(\vec n_r, \vec n_\theta, \vec n_\phi)
\vec\tau$ which satisfies $[\tau_i, \tau_j]=2i\ep_{ijk} \tau_k$ 
($i=r,\theta,\phi$),  the ansatz (\ref{YM-ansatz1}) is written as
\begin{eqnarray}
A^{(0)} =
\frac{1}{2e} \Big[   A_0\tau_r dt+A_1\tau_r
dr+(\phi_1\tau_{\theta}+(\phi_2-1)\tau_{\phi})d\theta \cr
\noalign{\kern 10pt}
 +(-(\phi_2-1)\tau_{\theta}+\phi_1\tau_{\phi})\sin\theta d\phi\Big]
\label{YM-ansatz2}
\end{eqnarray}
Note that there are no singularities in this gauge. Next make a gauge
transformation $A = S A^{(0)} S^{-1} - (i/e) dS \cdot S^{-1}$ where
\beeq
S=\left(
\matrix{+ e^{i(\phi+\Omega)/2} \cos \myfrac{\theta}{2} &
~~ +e^{-i(\phi-\Omega)/2} \sin \myfrac{\theta}{2} \cr
-e^{i(\phi-\Omega)/2} \sin \myfrac{\theta}{2} &
~~ +e^{-i(\phi+\Omega)/2} \cos \myfrac{\theta}{2} \cr}  \right)
~~,~~ \Omega=\Omega(t,r) ~~.
\label{gauge1}
\eneq
Useful identities are
\beqn
S \tau_r S^{-1} &=& \tau_3 \cr
S \tau_\theta S^{-1} &=& \cos \Omega \tau_1 - \sin \Omega \tau_2 \cr
S \tau_\phi S^{-1} &=& \sin \Omega \tau_1 + \cos \Omega \tau_2 \cr
2i dS \cdot S^{-1} &=&
- (\Omega' dr + \dot\Omega dt) \tau_3 
-  d\theta (\sin\Omega \tau_1 + \cos\Omega \tau_2)  \cr 
&& 
+ d\phi (\sin\theta \cos\Omega \tau_1 - \sin\theta \sin\Omega
\tau_2 - \cos\theta \tau_3)
\label{identity1}
\eeqn

The new gauge potential is 
\beeq
A=\frac{1}{2e} \, \Big\{ u\tau_3dt + \nu \tau_3 dr
+ (w\tau_1+\tilde{w}\tau_2)d\theta
+(\cot\theta\tau_3+w\tau_2-\tilde{w}\tau_1)\sin\theta d\phi\Big\} .
\label{YM-ansatz3}
\eneq
where
\beqn
u &=& A_0 + \dot\Omega \cr
\nu &=& A_1 + \Omega' \cr
w &=& +\phi_1 \cos\Omega + \phi_2 \sin\Omega \cr
\tilde w &=&  -\phi_1 \sin\Omega + \phi_2 \cos\Omega ~~~.
\label{YM-ansatz4}
\eeqn
Note that the gauge transformation (\ref{gauge1}) is singular at
$\theta=0$ and $\pi$.  Eq. (\ref{YM-ansatz4}) is the gauge potential
in the singular gauge.  It has a Dirac string.
 One can always choose $\Omega(t,r=0)=\pi/2$
with which the boundary conditions at $r=0$ are $u=\nu=\tilde w=0$ and
$w=1$.  With appropriate $\Omega(t,r)$ one can set
$\nu(t,r)=0$ or $u(t,r)=0$.

A straightforward calculation leads to  the field strength 
$F = dA-ieA\wedge A$:
\beqn
F&=&
\frac{1}{2e}\bigg\{
(\dot{\nu}-u^{\prime})\tau_3 dt\wedge dr
+\Big[ (\dot{w}-u\tilde{w})\tau_1 + (\dot{\tilde{w}}+uw)\tau_2\Big]
dt\wedge d\theta \cr
\noalign{\kern 8pt}
&&\hskip 1cm
-\Big[ (uw+\dot{\tilde{w}})\tau_1+(u\tilde{w}-\dot{w})\tau_2 \Big]
dt\wedge\sin\theta d\phi\cr
\noalign{\kern 8pt}
&&\hskip 1cm
+\Big[ (w^{\prime}-\nu\tilde{w})\tau_1
   +(\tilde{w}^{\prime}+w\nu)\tau_2\Big] dr\wedge d\theta \cr
\noalign{\kern 8pt}
&&\hskip 1cm
+ \Big[ (w^{\prime}-\nu\tilde{w})\tau_2 
  +(-\tilde{w}^{\prime}-\nu w)\tau_1\Big]   dr\wedge\sin\theta d\phi \cr
\noalign{\kern 8pt}
&&\hskip 1cm
-(1-w^2-\tilde{w}^2)\tau_3 d\theta\wedge\sin\theta d\phi \bigg\} ~~.
\label{field-strength1}
\eeqn
The configurations where $\nu=u=0$, $w=\tilde w=$constant, and
$w^2+\tilde w^2=1$ are pure gauge.

\subsection{Equations of motion}

In the general spherically symmetric metric (\ref{metric1})
tetrads are
\begin{equation}
e_0={\sqrt{H}\over p} dt ~~~,~~~
e_1={1\over \sqrt{H}}dr ~~~,~~~
e_2=rd\theta ~~~,~~~ e_3=r\sin\theta d\phi ~~~.
\label{tetrads}
\end{equation}
In the tetrad basis $F_{ab}=(e_a)_{\mu}(e_b)_{\nu}F^{\mu\nu}$ and
the energy-momentum tensors are
$T_{ab}= F_{ac}^{(i)}F_{b}^{c(i)} -
\frac{1}{4}\eta_{ab}F_{de}^{(i)}F^{de(i)}$.  The nonvanishing
components of the Yang-Mills equations (\ref{EYM-eq}) are
\beqn
&&\left(p r^2(u^{\prime}-\dot{\nu})\right)^{\prime} 
-2\frac{p}{H} \big\{ w(uw+\dot{\tilde{w}})
   +\tilde{w}(u\tilde{w}-\dot{w})\big\} =0 \cr
\noalign{\kern 10pt}
&&
\left( p r^2(u^{\prime}-\dot{\nu}) \right)_{,\;t}
-2\frac{H}{p} \big\{ -\tilde{w}w^{\prime}+\tilde{w}^{\prime}w
+\nu(w^2+\tilde{w}^2)\big\}= 0 \cr
\noalign{\kern 10pt}
&&
\left(\frac{H}{p} (w^{\prime}-\tilde{w}\nu) \right)^{\prime}
-\left({p\over H} (\dot{w}-u\tilde{w})\right)_{,\;t} \cr
\noalign{\kern 10pt}
&&
\hskip 2cm
+\frac{p}{H} u(uw+\dot{\tilde{w}})
+\frac{w(1-w^2-\tilde{w}^2)}{p r^2}
-\frac{H}{p}\nu(\tilde{w}^{\prime}+w\nu)  =0 \cr
\noalign{\kern 10pt}
&&
\left(\frac{H}{p} (\tilde{w}^{\prime}+w\nu)\right)^{\prime}
-\left({p\over H}(\dot{\tilde{w}}+uw)\right)_{,\;t} \cr
\noalign{\kern 10pt}
&&
\hskip 2cm
+\frac{p}{H} u(u\tilde{w}-\dot{w})
+ \frac{\tilde{w}(1-w^2-\tilde{w}^2)}{p r^2}
+\frac{H}{p}\nu(w^{\prime}-\tilde{w}\nu)
=0 ~~. \hskip 1cm
\label{YM-eq2}
\eeqn

The nonvanishing components of the energy-momentum tensor are given by
\beqn
T_{00} &=& {1\over e^2} ( A + B) \cr
\noalign{\kern 10pt}
T_{11} &=& {1\over e^2} (-A + B) \cr
\noalign{\kern 10pt}
T_{22} &=& T_{33} = {1\over e^2} A \cr
\noalign{\kern 10pt}
T_{01} &=& -{1\over e^2} \, C 
\label{energy-momentum1}
\eeqn
where
\beqn
A &=& {1 \over 2}\,  p^2 (\dot \nu - u')^2
  +  {1\over 2r^4} \,(1-w^2-\tilde{w}^2)^2 \cr
\noalign{\kern 10pt}
B &=& {p^2\over r^2 H} \Big\{  (uw+\dot{\tilde{w}})^2
   + (\dot{w}-u\tilde{w})^2 \Big\} 
+ {H\over r^2} \Big\{ (w^{\prime}-\nu\tilde{w})^2 
     + (\tilde{w}^{\prime}+\nu w)^2 \Big\} \cr
\noalign{\kern 10pt}
C &=& {2p\over  r^2} \Big\{
   (\tilde{w}^{\prime}+\nu w) (uw+\dot{\tilde{w}})
   + (w^{\prime}-\nu\tilde{w})(\dot{w}-u\tilde{w}) \Big\}  ~~.
\label{energy-momentum2}
\eeqn
The Einstein equations  reduce to 
\beqn
&&{p' \over p}  = - {8\pi G\over e^2} ~   {rB\over H} \cr
\noalign{\kern 10pt}
&&- {H'\over r} + {1-H\over r^2} = {8\pi G\over e^2} ~ (A+B) + \Lambda \cr
\noalign{\kern 10pt}
&&{p\over 2} \Bigg\{ \bigg( {p\dot H\over H^2} \bigg)_{,t}
 +  \bigg( {pH' - 2p'H\over p^2} \bigg)' \Bigg\} 
+{1-H\over r^2} = {16\pi G\over e^2} \, A \cr
\noalign{\kern 10pt}
&&{p\dot H\over rH} = - {8\pi G\over e^2} \, C
\label{Einstein2}
\eeqn
It is convenient to introduce  $m(r)$ defined by
\beeq
H(r) = 1 - {2m(r)\over r} - {\Lambda r^2\over 3} ~~.
\label{m-function}
\eneq
$m(r)$ is the mass contained inside the radius $r$.  $p(r)=$constant
and $m(r)=0$ corresponds to the Minkowski, de Sitter, or
anti-de Sitter space for $\Lambda=0, >0,$ or $<0$, respectively.
Then the second equation in (\ref{Einstein2}) becomes 
\beeq
m' = {4\pi G\over e^2} \, r^2 (A+B) ~~.
\label{Einstein3}
\eneq
The system of the Einstein-Yang-Mills equations contains one 
redundant equation.  The third equation in (\ref{Einstein2}) follows
from (\ref{YM-eq2}) and the rest of (\ref{Einstein2}).  


\subsection{Static configurations}

It is most convenient to take the $\nu=0$ gauge for static
configurations.  The second equation in (\ref{YM-eq2}) then yields
$w \tilde w' - w' \tilde w=0$, which leads to $\tilde w(r) = C w(r)$.
By a further global rotation $\Omega=$constant in (\ref{YM-ansatz4})
one can set $\tilde w =0$.  As a result 
\beqn
A&=&\frac{1}{2e} \Big\{ u\tau_3dt +w\tau_1d\theta
+(\cot\theta\tau_3+w\tau_2)\sin\theta d\phi\Big\} \cr
\noalign{\kern 10pt}
F&=&
\frac{1}{2e}\bigg\{
 -u^{\prime} \tau_3 dt\wedge dr
+ uw   dt\wedge (\tau_2 d\theta - \tau_1 \sin\theta d\phi) \cr
\noalign{\kern 8pt}
&&\hskip 1cm
+   w^{\prime}  dr\wedge (\tau_1 d\theta + \tau_2 \sin\theta d\phi)
-(1-w^2 )\tau_3 d\theta\wedge\sin\theta d\phi \bigg\} ~~.
\label{YM-static1}
\eeqn

Then the  Einstein-Yang-Mills equations are
\beqn
\left(\frac{H}{p}w^{\prime}\right)^{\prime} 
&=& -\frac{p}{H}u^2w-\frac{w}{p}\frac{(1-w^2)}{r^2} 
\label{YM1}  \\
\left(r^2pu^{\prime}\right)^{\prime}& =& \frac{2p}{H}w^2u  
\label{YM2}\\
p^{\prime} &=&
-\frac{2v}{r}p\left[(w^{\prime})^2+\frac{u^2w^2p^2}{H^2}\right] 
\label{Ein1}  \\
m' &=& v\left[
\frac{(w^2-1)^2}{2r^2}+\frac{1}{2}r^2p^2(u^{\prime})^2
  +H(w^{\prime})^2+\frac{u^2w^2p^2}{H} \right] 
\label{Ein2}
\eeqn
where  $v ={4\pi G}/{e^2}$. 
 
These equations are solved with the given boundary conditions.
Near the origin solutions must be regular so that
\begin{eqnarray}
u(r) &=& ar+\frac{a}{5}
  \big\{ -2b +\frac{1}{3}\Lambda + 2v(a^2+4b^2) \big\} r^3 \cr
w(r) &=& 1-br^2 \cr
m(r) &=& {1\over 2} v (a^2 + 4b^2)  r^3 \cr
p(r) &=& 1 - v (a^2 + 4b^2 ) r^2
\label{near-origin}
\end{eqnarray}
where $a$ and $b$ are arbitrary constants.
The boundary conditions  at the origin of the EYM equations are 
completely determined by the values of the constants $a$ and $b$.

At space infinity the energy-momentum tensors $T_{ab}$ in
(\ref{energy-momentum1}) must approach zero sufficiently fast.
Further we expect that the metric must asymptotically (anti-) de Sitter
space, depending on the value of $\Lambda$.  This, with the equations
of motion, leads to the asymptotic expansion at large $r$;
\begin{eqnarray}
u = u_0+u_1\frac{1}{r} + \cdots
~~~,~~~w = w_0+w_1\frac{1}{r} + \cdots\nonumber\\ 
m = M+m_1\frac{1}{r} + \cdots~~~,~~~
p = p_0+p_4\frac{1}{r^4} + \cdots
\label{INFTY}
\end{eqnarray}
where $u_0$, $u_1$, $w_0$, $w_1$, $m_1$, $p_0$ and $p_4$ are constants to
be determined and $M$ is  the ADM mass, $M=m(\infty)-m(0)$.

\section{Conserved charges} 

Solutions to eq.'s (\ref{YM1}) to (\ref{Ein2}) are classified by the
ADM mass, $M=m(\infty)-m(0)$, electric and
magnetic charges, $Q_E$ and $Q_M$.  From the Gauss flux theorem 
\begin{equation}
\pmatrix{Q_E\cr Q_M\cr}
 = {e\over 4\pi} \int dS_k \, \sqrt{-g} \, 
\pmatrix{ F^{k0}\cr \tilde F^{k0} \cr}
\label{charge1}
\end{equation}
are conserved.  With  the ansatz in the
singular gauge (\ref{YM-ansatz3}) and the asympotitic behavior
(\ref{INFTY}), the charges are given by
\begin{equation}
\pmatrix{Q_E\cr Q_M\cr}
 = \pmatrix{u_1 p_0\cr 1 - w_0^2\cr} {\tau_3\over 2}
\label{charge2}
\end{equation}
Notice that the electric charge $Q_E$ is determined by $u_1$, whereas
the magnetic charge $Q_M$ by $w_0$.  If $(u,w,m,p)$ is a solution,
then $(-u,w,m,p)$ is also a solution.  Dyon solutions come in a pair
with $(\pm Q_E, Q_M, M)$. 

The charges (\ref{charge1}) are not gauge invariant, however.  
Under a local gauge transformation 
$A \go U A U^{-1} - (i/e) dU U^{-1}$, $Q_E$ and $Q_M$
are  transformed to
\begin{equation}
\pmatrix{{Q_E}^U\cr {Q_M}^U\cr}
 = {e\over 4\pi} \int dS_k \, \sqrt{-g} \, U(x)
\pmatrix{ F^{k0}\cr \tilde F^{k0} \cr} U^{-1}(x)
\label{charge3}
\end{equation}
In non-Abelian gauge theory a set of charges $\{ {Q_E}^U ,
{Q_M}^U \}$ are conserved.  In the rest of the paper we use the 
charges, (\ref{charge2}), defined in the singular gauge.

The effective charge $Q_{\rm eff}$ \cite{BARTNIK} is defined by 
the asymptotic behavior of $H(r)$;
\beeq
H(r) = 1-\frac{2M}{r}+\frac{Q_{\rm eff}^2}{r^2}
-\frac{1}{3}\Lambda r^2 ~~~.
\label{eff-charge}
\eneq
In terms of the coefficients in (\ref{INFTY}),
$Q_{\rm eff}^2 = -2m_1$. This requires that $m_1<0$  which indeed is the
case. After inserting eq. (\ref{INFTY})
into eq.'s (\ref{YM1}) and (\ref{Ein2}) we find the relation
\begin{equation}
Q_{\rm eff}^2 = 
2 v {\rm Tr} (Q_E^2+Q_M^2) -  \frac{4\Lambda}{3}\frac{p_4}{p_0}.
\label{CHARGE_REL}
\end{equation}
Eq.\ (\ref{Ein1}) implies that $p(r)$ is a monotonically decreasing
positive function so that $p_0>0$ and $p_4>0$.  The effective charge is
smaller (larger) than $2 v {\rm Tr} (Q_E^2+Q_M^2)$ for $\Lambda >0$ 
($<0$).  The relation (\ref{CHARGE_REL}) incidentally implies that
the charges defined in the singular gauge have physical, gauge
invariant meaning.

\section{Sum rules}

Sum rules are obtained from the equations of motion.
First, multiply both sides of (\ref{YM2})  by $u$ and integrate in part.
\begin{equation}
 pr^2u u^{\prime}\Bigg|_{r_1}^{r_2} =
\int_{r_1}^{r_2} dr \, \Bigg\{ r^2 p (u^{\prime})^2  +
2\frac{p}{H}u^2w^2 \Bigg\}  ~~.
\label{sum-rule1}
\end{equation}
Secondly, multiply  both sides of (\ref{YM1}) by $w$ and
integrate in part:
\begin{equation}
 \frac{H}{p} ww^{\prime} \Bigg|_{r_1}^{r_2} 
= \int_{r_1}^{r_2}  dr \, 
\Bigg\{ \frac{H}{p}(w^{\prime})^2 -\frac{p}{H}u^2 w^2
  -\frac{1}{pr^2} w^2(1-w^2) \Bigg\} ~~.
\label{sum-rule2}
\end{equation}
Thirdly, divide both sides of (\ref{YM1}) by $w$ and
integrate in part:
\begin{equation}
- \frac{Hw^{\prime}}{pw} \Bigg|_{r_1}^{r_2}  
= \int_{r_1}^{r_2} dr \, \Bigg\{  \frac{p}{H}u^2 +
\frac{H}{p}\left(\frac{w^{\prime}}{w}\right)^2
+\frac{1}{pr^2}(1-w^2) \Bigg\} ~~.
\label{sum-rule3}
\end{equation}
These relations are valid, provided the integrals on the right
hand sides are defined.  Several important conclusions follow from
(\ref{sum-rule1}) -  (\ref{sum-rule3}).

\subsection{In asymptotically flat space}
Consider (\ref{sum-rule1}) with $r_1=0$ and $r_2=\infty$. For regular
solutions $u(0)=0$. Both $p$ and $H$ approach constant as $r \go \infty$.
The finiteness of the ADM mass requires that 
$u w \big|_{r=\infty}= 0$.  In the expansion (\ref{INFTY}), $u_0 w_0=0$.
On the other hand, if $u_0 \not= 0$, $w_0=0$ and Eq.\ (\ref{YM2}) implies
$w_1\not= 0$ so that  Eq.\ (\ref{YM1}) cannot be satisfied.  Hence
$u(\infty)=0$.  Then the left hand side of (\ref{sum-rule1}) vanishes,
implying that $u(r)$ must vanish identically.  There is no regular
electrically charged solution. Furthermore,  (\ref{YM1}) can be solved only if $(w_0)^2 =1$ as
$H(\infty)=1$, therefore the magnetic charge $Q_M$ vanishes.  

Suppose that $w(r)$  never vanishes and $w^2 \le 1$ for 
$0<r<\infty$.  Consider (\ref{sum-rule3}) with $r_1=0$ and $r_2=\infty$.
The l.h.s. vanishes, but the integrand on the r.h.s.\ is 
positive definite except for the pure gauge configuration $w(r)=\pm 1$. 
This implies that non-trivial solutions with $w^2 \le 1$ must vanish
at least once.

We also note that the singular 
solution $w(r)=0$, $u'(r)=r^{-2}$, and $p(r)=1$ is nothing but the 
Reissner-Nordstr\"om solution.

\subsection{In asymptotically de Sitter space}
In asymptotically de Sitter space $H(r) \rightarrow -\Lambda r^2/3$ as
$r\rightarrow\infty$. In this case the finiteness of the ADM mass
does not forbid non-vanishing $uw$ at $r=\infty$.  However, there arises
a cosmological horizon at $r=r_h$ where $H(r_h)=0$.  

It follows from (\ref{Ein1}) and (\ref{Ein2}) that $u$ or $w$ must vanish
at $r=r_h$.  Now consider (\ref{sum-rule3}).  Suppose that $w(r_h) \not=
0$, or equivalently $w \not= 0$ for $r_1 \le r \le r_2=r_h$ for some 
$r_1>0$,  the left hand side is finite so that $u(r_h)=0$.  However,
Eq.\ (\ref{YM2}) implies that $u(r)$ is a monotonically increasing
or decreasing function.  With the boundary condition $u(0)=0$,
the only possibility available is $u(r)=0$.  This conclusion remains valid
even if $w(r_h)=0$.  In this case the first and second terms on the
r.h.s.\ give positively divergent contributions near $r_h$, whereas 
the l.h.s.\ remains finite.  To summarize, there is no solution with
nonvanishing $u(r)$.

If $w >0$ for  $0\le r \le r_h$, then the left hand side vanishes in the 
$r_1\go 0$ and $r_2 \go r_h$  limit. If one further assumes that
$w^2 \le 1$ in the interval, the integrand on the right hand side
is positive definite so that the only solution is 
$w(r) = \pm 1$, which is a pure gauge.  

This argument also shows that a nontrivial solution $w(r)$, which 
satisfies $w^2 \le 1$ for $0 \le r \le r_h$, must vanish at least once
in this interval.  We have numerically looked for  solutions in which $w^2
\ge 1$ for small $r$, but have found that no such solution exists.

\subsection{In asymptotically anti-de Sitter space}
In asymptotically anti-de Sitter space there exist solutions in
which $H(r) > 0$ everywhere.  As $H(r) \sim |\Lambda| r^2/3$
for large $r$, the condition for the finiteness of the ADM mass 
requires only that $uw \go$ constant.  In other words, both $u$ and
$w$ may approach nonvanishing values as $r\go\infty$.  Consequently,
with the expansion (\ref{INFTY}) the l.h.s. of (\ref{sum-rule1})
with $r_1=0$ and $r_2=\infty$  is $- p_0 u_0 u_1$ and can be
non-vanishing.   Solutions with $u(r)\not=0$ are allowed.

In critical cases a cosmological horizon appears and $H(r_h)=0$.
In this case the argument above for the asymptotically de Sitter
case applies and either $u$ or $w$ must vanish at $r=r_h$.  
We shall find, indeed,    $w(r_h)=0$ below.

\section{Soliton solutions}

\subsection{The BK solution}

Particle like solutions of the EYM equations in asymptotically
Minkowski space were first found by Bartnik and McKinnon (BK)
\cite{BARTNIK} in 1988. When $u=0$, we can solve the equations
(\ref{YM1}) to (\ref{Ein2}) numerically with the boundary condition $a=0$.

As already discussed, $w(r)=1$ and $H(r)=1$ corresponds to a pure
gauge configuration.   Similarly, if $w =0$ as $r\rightarrow\infty$ we are
left with the RN solution which is singular at $r=0$. 
The Yang-Mills charge (\ref{charge2})
vanishes in asymptotically Minkowski space since $w_0 = \pm 1$.
The effective charge $Q_{\rm eff}=0$ in all BK solutions, since 
$Q_M = \Lambda =0$.  All solutions   possess a metric that is
asymptotically Schwarzschild.

There is a discrete set of BK solutions, labeled by
the number of nodes in $w$, $n \in [1,\infty)$, and the free shooting parameter
$b$.  Since all BK solutions have at least one node, $n\ge 1$, the
solutions are unstable against spherically symmetric perturbations \cite{ZHOU}.

\subsection{Solutions with a cosmological horizon}

Solutions in asymptotically de Sitter space were obtained by
adding a cosmological constant ($\Lambda$) term
to the Einstein equations. Solutions display the same basic properties
as the BK  solutions\cite{VOLKOV}.  Just as in the BK solution, these
equations are solved numerically, using the shooting method and  
requiring   $a=0$.

In asymptotically de Sitter space, a cosmological horizon,  where $H=0$, 
develops at $r=r_h$. At the horizon $H'(r_h) \not= 0$.   
Near the horizon
\beqn
w(r) &=& w_0 + w_1 x + w_2 x^2 + \cdots \cr
p(r) &=& p_0 + p_1 x + p_2 x^2 + \cdots \cr
m(r) &=& m_0 + m_1 x + m_2 x^2 + \cdots \cr
H(r) &=&  h_1 x + h_2 x^2 + \cdots 
\label{horizon1}
\eeqn
where $x = r-r_h$.  $m_0$ and $r_h$ are related by
$1 - (2m_0/r_h) - (\Lambda r_h^2/3) =0$.  With given $w_0$, $r_h$,
and $p_0$, the equations (\ref{YM1}), (\ref{Ein1}), and (\ref{Ein2})
(with $u=0$) determine all other coefficients, provided
$w_0\not= 0, \pm 1$ and $p_0\not= 0$;
\beqn
m_1 &=& {v\over 2 r_h^2}  ~(w_0^2 - 1)^2 \cr
\noalign{\kern 8pt}
h_1 &=& -{2m_1\over r_h} + {2m_0\over r_h^2} - {2\over 3} \Lambda r_h \cr
\noalign{\kern 8pt}
w_1 &=& - {w_0 (1-w_0^2)\over h_1 r_h^2} \cr
\noalign{\kern 8pt}
p_1 &=& - {2v w_1^2 p_0\over r_h}
\label{horizon2}
\eeqn
and so on.   This expansion is valid
independent of the value of $\Lambda$.  The critical case 
$h_1 = H'(r_h) =0$ requires a special treatment, and will be 
analyzed in Section (5.3).  K\"unzle and Masood-ul-Alam \cite{KUNZLE}
have argued that $w(r)$ has $\sqrt{|r - r_h|}$ singularity at
$r=r_h$.  However, we have found that the regular expansion
(\ref{horizon1}) is valid.  

Just like the BK case, there are a discrete set of solutions labeled
by the number of nodes in $w(r)$, $n$ and the parameter $b$. Solutions
in $w$ and $m$, have the same form as the BK solutions except that
$w(r)$ no longer approaches 1 at infinity. Eq. (\ref{CHARGE_REL})
implies that there is a non-vanishing charge $Q_M$ (or $w \ne 1$) if
$\Lambda \ne 0$ and that the solutions are all asymptotically 
Reissner-Nordstr\"om type.
$|w(\infty)|$ is slightly greater than 1 for the $n=1$ solutions. When
$n \ge 2$, $1> w(\infty) > 0$ where $w(\infty) \rightarrow 0$ as
$n\rightarrow\infty$. 
 
The mass $m(r)$ also stays finite. 
For all indices $n$, it is small near the
origin and does not grow until it approaches the horizon where it
quickly climbs to a value near 1. After the horizon it stays almost
constant.

The position of the horizon  depends on
$\Lambda$. As long $\Lambda$ is below some critical value,  
the geometry approaches Reissner-Nordstr\"om-de Sitter space in the
asymptotic region as indicated by Eq.\ (\ref{CHARGE_REL}).
Above some critical value the topology changes and
a singularity appears. Since the position of the horizon depends
on $\Lambda$, there is a value for $\Lambda$ where this singularity
and the horizon meet, in which the topology becomes a completely
regular manifold. Above this value for $\Lambda$, the solutions are no
longer regular. More details of the topology dependence on $\Lambda$
can be found in \cite{VOLKOV}. All the  solutions are
unstable \cite{BRODBECK}.

\subsection{Solutions in asymptotically anti-de Sitter space}

As already discussed, there are no boundary conditions that forbid a
solution to the EYM equations in asymptotically anti-de Sitter space with a
non-zero electric component, $u(r)$, to the Yang-Mills fields.  Solutions
to Eqs.\ (\ref{YM1}) to (\ref{Ein2}) are determined with the cosmological
constant
$\Lambda$ fixed at some negative value.

\leftline{\bf (a) Monopole solutions}

Monopole solutions are obtained by setting $a=0$ 
($u=0$). By varying the initial condition parameter $b$, a continuum of
monopole solutions are found which are regular in the entire
space.  Just as in the BK and dS solutions, $w$ crosses the axis an
arbitrary number of times depending on
the value of the adjustable shooting parameter $b$. In contrast to
the  $\Lambda=0$ and $\Lambda>0$ cases which have a discrete set of
solutions in $b$ and
$n$,  there is a continuum of solutions in $b$ for each $n$.
Typical solutions are displayed  in fig.\ 1.

\begin{figure}[tbh]\centering
 \leavevmode 
\mbox{
\epsfysize=9.0cm \epsfbox{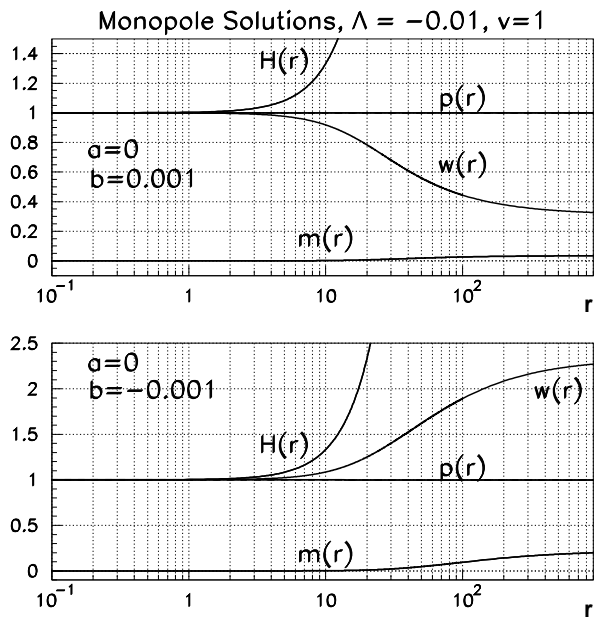}}
\caption{Monopole solutions for  $\Lambda = -0.01$ and $v=1$.
$(a,b)=(0, 0.001)$ and $(0, -0.001)$.  
$(w,m)$ at $r=\infty$ are $(0.318, 0.035)$ and $(2.304, 0.201)$. }
\label{m_plot}
\end{figure}

The behavior of $m$ and $p$ is similar to that of the asymptotically dS
solutions \cite{VOLKOV}. In contrast, as
shown in fig.\ \ref{m_plot}, there exist solutions where $w$ has no
nodes. These solutions are of particular interest because they are shown
to be stable against linear perturbations.

\leftline{\bf (b) Dyon solutions}

\begin{figure}[bth] \centering
\leavevmode 
\mbox{
\epsfxsize=9.0cm \epsfbox{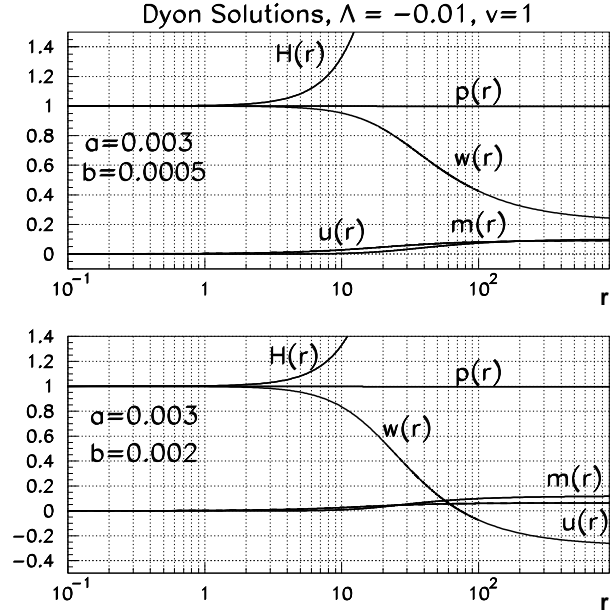}}
\caption{Dyon solutions for  $\Lambda = -0.01$ and $v=1$. In the top
figure $w(r)$ has no node ($n=0$), whereas in the bottom figure it has
one node ($n=1$).}
\vskip -0cm
\label{d_plot}
\end{figure}

Dyon solutions to the EYM equations are determined if
the adjustable shooting parameter $a$ is chosen to be non-zero
for a given negative $\Lambda$. Just as in the monopole solutions, 
 we find a continuum of solutions where $w$ crosses the axis an
arbitrary number of times depending on $a$ and $b$. Also similar to
the monopole solutions is the existence of solutions where $w$ does not
cross the axis.  As shown in Fig. \ref{d_plot}, the electric component, $u$, of the EYM
equations starts at zero and monotonically increases to some finite
value. The behavior of $w$, $m$ , $H$, and $p$ is similar to that in the
monopole solutions.

Just as for the monopole case, dyon solutions are found for a continuous
set of parameters,  $a$ and $b$.  For some values of $a$ and
$b$,  solutions blow up, or  the function $H(r)$ crosses the
axis and becomes negative.

\ignore{
We do not expect
there to be dyon solutions in this case, since there can be no
electric component to the EYM equations in the presence of a
cosmological horizon.}

\leftline{\bf (c) Critical solutions}

As the parameter $b$ is increased, the minimum of $H(r)$ hits 
zero from above, i.e.\ $H(r_h) =H'(r_h)=0$.  This constitutes
a special case and needs careful examination.     Numerical studies
indicate that this happens in a finite range of the parameter $a$.
The critical solution exists for both $u(r)=0$ and $u(r) \not= 0$ cases.
One example of solutions near the critical value
[$(a,b)=(0.01, 0.69)$] is displayed in fig. 3.  $H(r)$ becomes very close
to zero at $r \sim 1$. It has $(Q_E, Q_M, M) \sim (0.015,0.998,0.995)$.

When $b=b_c$, $w$ and $p$ vanish at $r=r_h$ as well.   As $p(r_h)=0$,
$p(r)=0$ for $r \ge r_h$.  The space ends at $r=r_h$.  There is 
universality at the critical point.

\begin{figure}[thb]\centering
\leavevmode 
\mbox{
\epsfysize=7.cm \epsfbox{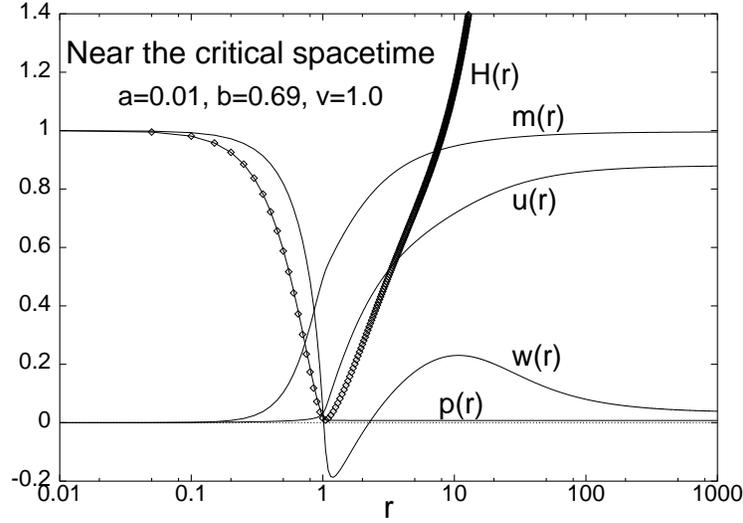}}
\caption{Dyon solution for  $\Lambda = -0.01$, $v=1$, $a=0.01$ and 
$b=0.69$.  At $a=0.01$,  the critical value is $b_c=0.7104$.  $H$
almost hits the axis around $r=1$.  At $b_c$, $p(r)=0$ for $r\ge r_h$.
The space ends at $r_h$.}
\label{critical_d_plot}
\end{figure}

The numerical integration of the differential equations indicates
that $m(r)$ and $u(r)$ are regular at $r=r_h$.  The appropriate
ansatz for the critical solutions with $H(r_h)=H'(r_h)=0$ is,
for $y=r_h - r \ge 0$,
\beqn
u &=&  u_0 + u_1 y + u_2 y^2 + \cdots \cr
w &=& y^\alpha \Big\{ w_0 + w_1 y + w_2 y^2 + \cdots \Big\} \cr
p &=& y^\beta \Big\{ p_0 + p_1 y + p_2 y^2 + \cdots \Big\} \cr
H &=& y^\gamma \Big\{ h_0 + h_1 y + h_2 y^2 + \cdots \Big\} \cr
m &=&  m_0 + m_1 y + m_2 y^2 + \cdots  
\label{horizon3}
\eeqn
Eq.\ (\ref{YM1}) implies that $\gamma=2$.  When $u(r)\not= 0$,
Eq.\ (\ref{YM2}) leads to $\alpha=\onehalf$.  When $u(r)=0$,
Eq.\ (\ref{Ein1}), instead,  implies that $\alpha=\onehalf$.
Other relations obtained from eqs.\ (\ref{YM1}) - (\ref{Ein2})
are
\beqn
\Big(\beta - {3\over 2} \Big) h_0 &=& {2\over r_h^2} \cr
\noalign{\kern 8pt}
\beta r_h^2 u_1 &=& {2w_0^2 u_0\over h_0} \cr
\noalign{\kern 8pt}
\beta &=& {v w_0^2\over 2 r_h} \cr
\noalign{\kern 8pt}
m_1 &=& - {v\over 2 r_h^2} ~~~.
\label{horizon4}
\eeqn
The value of the index $\beta$ is unconstrained when $u=0$.
However, if $u\not= 0$, the consistency of eq.\ (\ref{YM1}),
for instance, demands that $2\beta$ be an integer.   The smallest
value for $\beta$ which satisfies the first relation in (\ref{horizon4})
is $\beta=2$, as $h_0>0$.  We have confirmed   this by numerical
studies.   The relation (\ref{m-function}) further implies that
\beqn
{2m_0\over r_h} &=&1 -  {\Lambda \over 3} r_h^2 \cr
\noalign{\kern 8pt}
m_1 &=& {1\over 2} ( \Lambda r_h^2 - 1) ~.
\label{horizon5}
\eeqn
From the two relations for $m_1$, one in (\ref{horizon4}) and the other 
in (\ref{horizon5}),   $r_h$ is determined as a function of $v$ and
$\Lambda <0$:
\beeq
r_h^2 = {1\over 2|\Lambda|} 
\left( \sqrt{1 + 4v|\Lambda|} - 1 \right) ~~.
\label{critical-horizon1}
\eneq

To summarize, the indices in (\ref{horizon3}) are given by
\beeq
\alpha = {1\over 2} ~~,~~ \beta =\gamma = 2 ~~.
\label{universality1}
\eneq
The coefficients $m_0$, $m_1$, and 
\beeq
w_0^2 = {4r_h\over v} ~~,~~ h_0 = {4\over r_h^2} 
\label{universality2}
\eneq
are all determined by $v$ and $\Lambda$ only.  We are observing
the universality in the behavior of the critical solutions.  The
coefficients  $u_0$ and $p_0$ depend on $a$ or $b$ as well.

For small $v|\Lambda| \ll 1$
\beqn
&&r_h \sim \sqrt{v}  ~~,~~
m_0 \sim {1\over 2} \sqrt{v} ~~,~~
m_1 \sim - {1\over 2} \cr
\noalign{\kern 10pt}
&&w_0 \sim  {2 \over v^{1/4}}  ~~,~~  
h_0 \sim  {4\over v}
\label{universality3}
\eeqn
They are all determined by $v$ only.
This universal behavior is clearly seen in the solution in fig. 3 which
is very close to the critical one.

The meaning of the critical spacetime is yet to be clarified.  The
space ends at $r=r_h$.  It defines a spacetime with a boundary.

\begin{figure}[tbh]\centering
\leavevmode 
\mbox{
\epsfysize=7.0cm \epsfbox{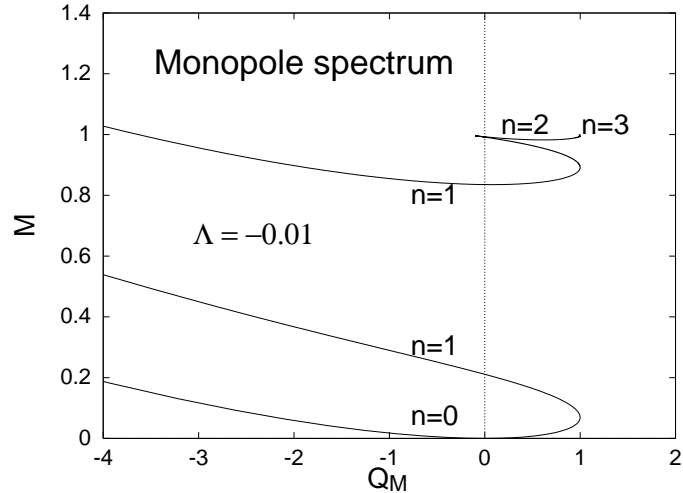}}
\caption{Mass $M$ is plotted as a function of magnetic charge $Q_M$
for monopole solutions at $\Lambda=-0.01$ and $v=1$.  The number of nodes,
$n$,  in $w(r)$ is also  marked.  In the lower branch $Q_M=1$ at 
$b=0.00168$.}
\label{qm_plot}
\end{figure}

\bigskip
\leftline{\bf (d) Spectrum of monopole and dyon solutions}
Monopole and dyon solutions permit non-vanishing charges $Q_M$ and
$Q_E$, although there are solutions where $Q_M =0$ and $Q_E\ne 0$ or
where $Q_E=0$ but $Q_M\ne 0$. Non-zero charges $Q_M$ or $Q_E$ ensures
that $Q_{\rm eff}\ne 0$ (see Eq. (\ref{CHARGE_REL}) ) so that solutions
are asymptotically of the AdS Reissner-Nordstr\"om type.

In fig.\ \ref{qm_plot} the mass $M$ is plotted as a function of $Q_M$ for
monopole solutions at $\Lambda=-0.01$ and $v=1$. The behavior of the
solutions near t$b=b_c=0.7104$ needs more careful analysis.

\begin{figure}[tbh]
\centering \leavevmode 
\mbox{
\epsfysize=7.cm \epsfbox{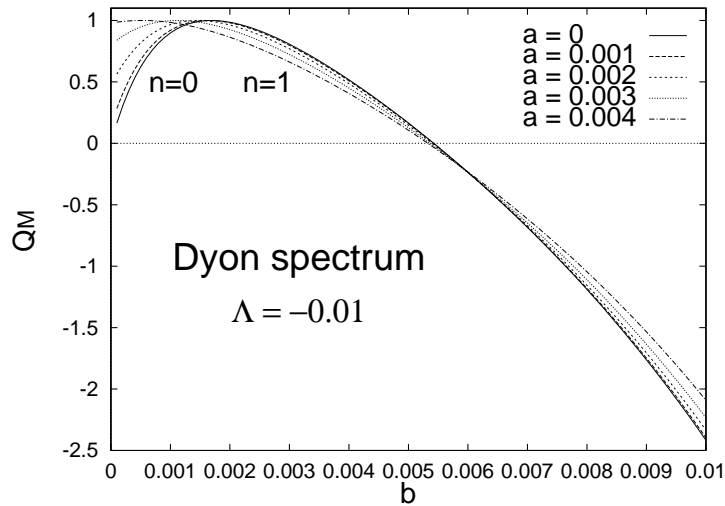}}
\caption{Magnetic charge $Q_M$ of dyon solutions as a function of the 
parameter $b$ is plotted with various values of the parameter $a$.
At $b \sim 0.0061$, $Q_M$ is independent of $a$,
taking the quantized value $-(4\pi)^{-1/2}=-0.2821$ within  
numerical errors.}
\label{BM_SPECTRUM}
\end{figure}

Dyon solutions are found in a  good portion of the $Q_E$-$Q_M$
plane.  There are solutions with $Q_M=0$ but $Q_E\not=0$. 
Although $Q_M=0$, i.e. $w(\infty)=\pm 1$, $w(r) \not= 0$.  In the shooting
parameter space $(a,b)$,  these solutions correspond not exactly, but
almost to a universal value for $b \sim 0.0054$.  See fig.\
\ref{BM_SPECTRUM}.   More surprising is the fact that 
$Q_M$ takes a quantized value $  - (4\pi)^{-1/2}$ at $b=0.0061$
independent of the value of $a$ within numerical errors.  We have not
understood why it should be so.

\begin{figure}[bth]
\centering \leavevmode 
\mbox{
\epsfxsize=8.0cm \epsfbox{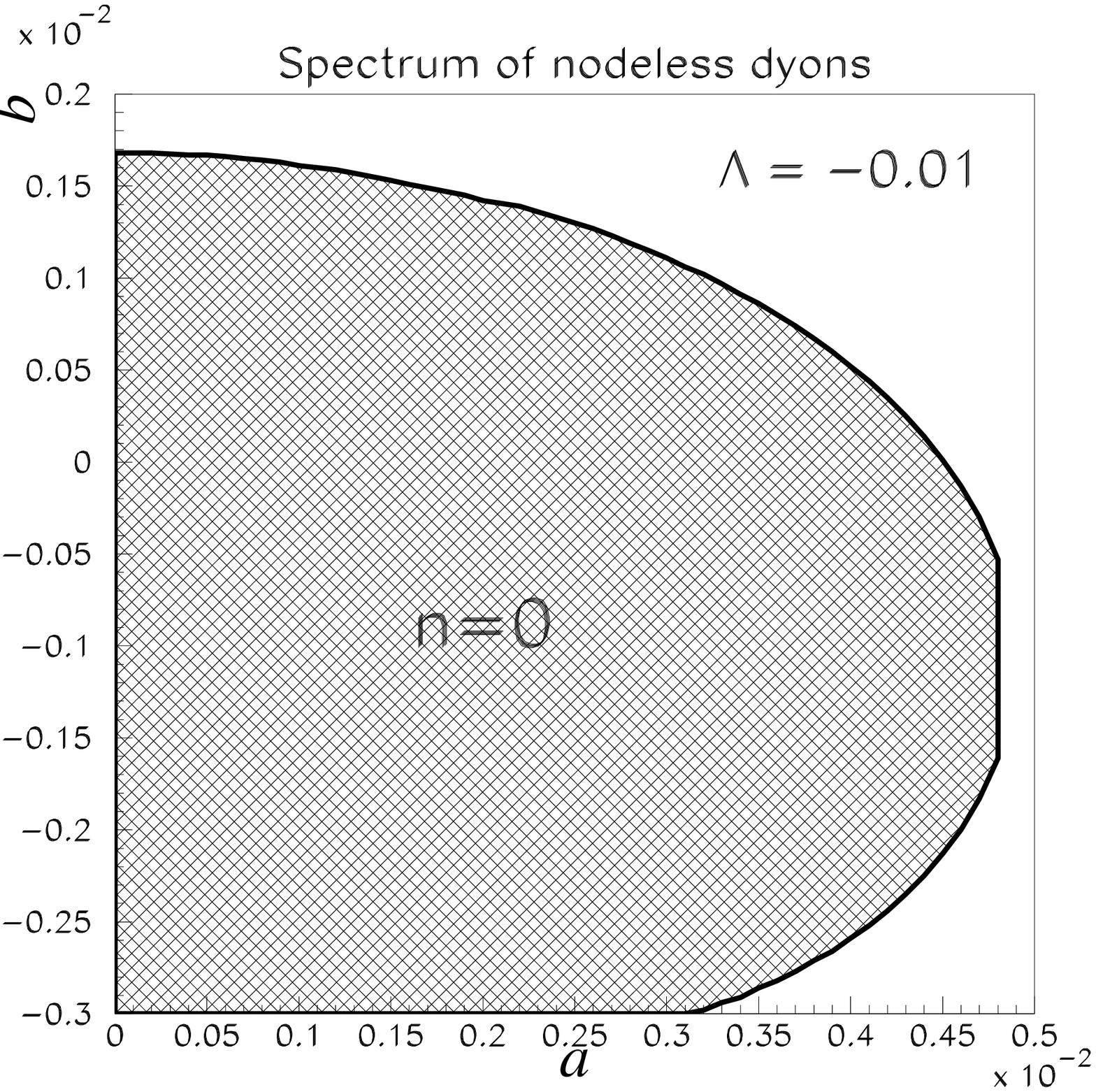}}
\caption{Spectrum of nodeless dyons.}
\label{dyon-spectrum1}
\vskip -0cm
\end{figure}

Solutions with no node in $w(r)$ have special importance,
as they are stable against small fluctuations.  (See section 8.)
In fig.\ \ref{dyon-spectrum1} the spectrum of nodeless dyon solutions
are presented in the parameter space $(a,b)$.  Notice 
that $a$ must be small enough ($a<0.005$) even for $b<0$.

\bigskip

\leftline{\bf (e) Dependence of the coupling $v$ on the solutions}

\begin{figure}[tbh]\centering
\leavevmode 
\mbox{
\epsfxsize=8.cm \epsfbox{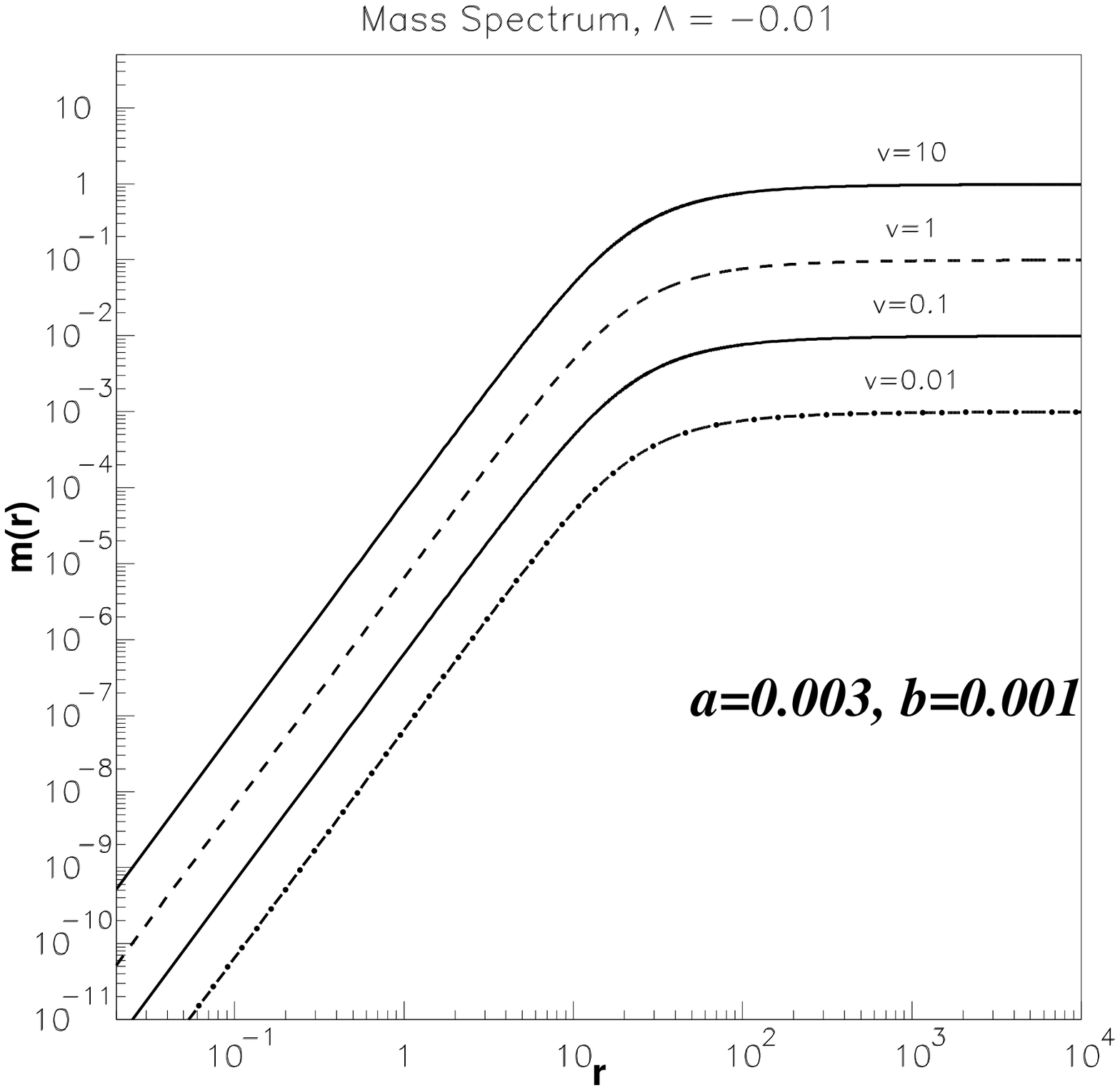}}
\caption{ A typical solution showing the dependence of $v$ on the mass
$m(r)$. $\Lambda = -0.01$, $a=0.003$ and $b=0.001$.  $w(r)$ and $u(r)$
do not have much dependence on $v$.}
\label{V_MASS}
\end{figure}

The ADM mass $M$ depends on the value of the coupling $v=4\pi G/e^2$. 
As shown in fig.\ \ref{V_MASS}, $M$ increases as
$v$ gets larger and decreases when $v$ gets smaller. 
With fixed $(a,b)$, roughly $M \propto v$.  The Yang-Mills
fields $u$ and $w$ are roughly independent of $v$.

\section{Black hole solutions}

Not long after the BK solutions were discovered, black hole solutions
were also found to be contained in the EYM
equations\cite{BIZON} if different boundary conditions were
used.  These solutions generated a large amount of
further study, as they apparently violate the no-hair conjecture
\cite{HEUSLER}.  Later, EYM black holes were studied in a cosmological
context by including a positive cosmological constant\cite{VOLKOV}.
These black hole solutions share most of the properties as the  soliton
solutions, including their instability.

Recently, purely magnetic black hole solutions were found in
asymptotically anti-de Sitter space \cite{WINSTANLEY}.  These solutions
are drastically different  from their asymptotically Minkowski or de
Sitter counterparts. There   are  a
continuum of solutions in terms of the adjustable shooting parameter
that specifies the initial conditions at the horizon. Furthermore,
there exist solutions that have no node in $w$ and are in turn stable
against spherically symmetric linear perturbations.  Here, we discuss the
solutions found by Winstanley \cite{WINSTANLEY} and also present new dyon
black hole solutions.  We also discuss the apparent shrinking of the
moduli space when the magnitude of $\Lambda$ is decreased. Similar to the
particle-like solutions already discussed, the moduli space becomes
discrete in the 
$\Lambda\rightarrow 0$ limit.

\subsection{Boundary conditions at the horizon}

Black hole solutions are obtained numerically by specifying the
boundary conditions at the horizon and shooting for regular solutions
$w$, $u$, $m$ and $p$ for $r_h \le r < \infty$. The location of the
horizon, $r_h$, and the  value $p(r_h) > 0$ can be arbitrarily chosen by
scaling of $t$ and $r$.  We look for solutions in which $H(r) >0$ for
$r> r_h$.  As $H(r_h)=0$ but $p(r_h) \not= 0$, Eqs.\
(\ref{YM1}) - (\ref{Ein2}) require that either  $u$ or $w$ vanishes
at the horizon.  A stronger condition is obtained from the sum rule 
(\ref{sum-rule3}) with  $r_1=r_h$ and $r_2=\infty$. Its l.h.s.\  is
finite so that  $u(r_h)=0$ on its r.h.s.  Hence we are led to the 
expansion
\beqn
w &=& w_0 +w_1x + \cdots \cr
u &=& u_1 x + \cdots \cr
p &=& 1 + p_1x + \cdots \cr
H &=& h_1x + \cdots \cr
m &=& m_0 + m_1 x + \cdots 
\label{BH-expansion1}
\eeqn
where $x = r-r_h$. We have chosen $p(r_h)=1$ without loss of generality.  

There are two adjustable shooting parameters, $(a,b)=(u_1,w_0)$.
After inserting the ansatz into eq.'s (\ref{m-function}) to (\ref{Ein2}) 
we find
\beqn
m_0 &=& \frac{r_h}{2}-\frac{\Lambda r_h^3}{6} \cr
\noalign{\kern 8pt}
m_1 &=& {v\over 2} \Bigg\{ {(1-w_0^2)^2\over r_h^2} + u_1^2 \Bigg\} \cr
\noalign{\kern 8pt}
h_1 &=& \frac{1}{r_h} ( 1 -\Lambda r_h^2 - 2m_1) \cr
\noalign{\kern 8pt}
w_1 &=& -\frac{w_0(1-w_0^2)}{r_h^2 h_1} \cr
\noalign{\kern 8pt}
p_1 &=& - \frac{2v}{r_h} \Bigg\{ (w_1)^2
    +\frac{w_0^2 u_1^2}{h_1^2}\Bigg\} \cr
\noalign{\kern 8pt}
u_2 &=& - \Bigg\{ {w_0^2\over r_h^2 h_1}
       + {1\over r_h} + {p_1\over 2}  \Bigg\} ~~.
\label{BH-expansion2}
\eeqn
The asymptotic expansion at large $r$ is the same as in   (\ref{INFTY}).
The ADM mass is given by $M = m(\infty)$.

\subsection{New electrically and magnetically charged black hole
solutions}

\begin{figure}[htb] \centering
\leavevmode 
\mbox{
\epsfxsize=8.cm \epsfbox{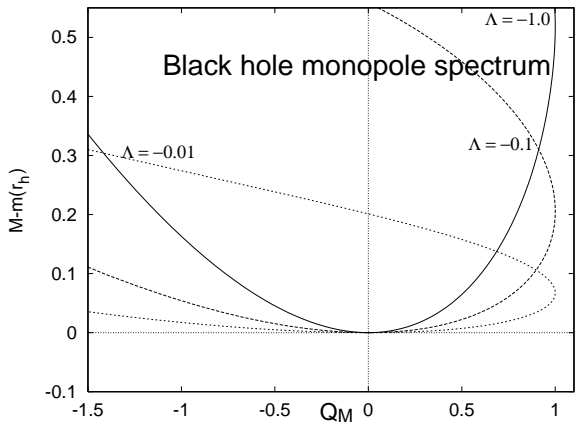}}
\caption{Black hole mass vs. magnetic charge $Q_M$ for the $n=0$ and
$n=1$ arms and for different values of $\Lambda$. }
\label{BH-spectrum1}
\end{figure}

Just as the  soliton solutions, purely magnetically charged black hole
solutions are obtained by setting the adjustable parameter $a$ to
zero. The behavior of the solutions are   similar to that of the solitons
 (see Ref. \cite{WINSTANLEY} for more information). 
The number of nodes $n$ in $w$ can be 0, 1, 2, $\cdots$.
The black hole monopole spectrum of mass versus charge is displayed in
fig.\ \ref{BH-spectrum1}.  It shows the spectrum for the
$n=0$ and $n=1$ arms. 

\begin{figure}[hbt]\centering
 \leavevmode 
\mbox{
\epsfxsize=8.cm \epsfbox{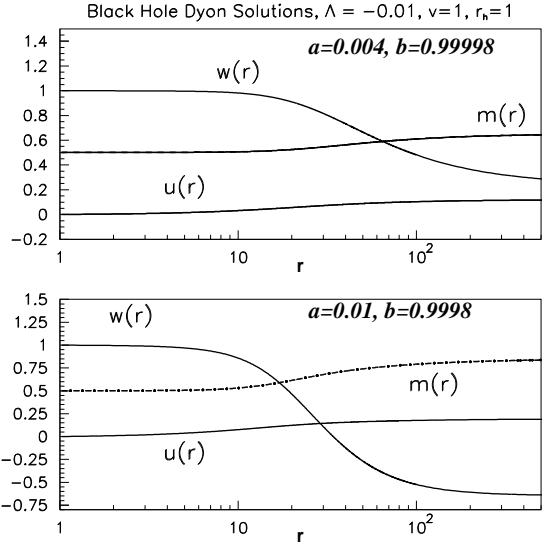}}
\caption{Typical dyon black hole solutions with no node and one node. 
$H(r)$ and $p(r)$ were not plotted for the sake of clarity. The top
figure corresponds to $(a,b)=(0.004, 0.99998)$, whereas the bottom
to $(0.01, 0.9998)$.}
\label{BH_dyons}
\end{figure}

Solutions with both magnetic and electric charge are
obtained by giving $a$ a finite value. Dyon black hole solutions
are similar to the monopole solutions except that $u$ is nonzero. At
the horizon $u$ starts at zero and monotonically increases
asymptotically to a finite value. $H$ starts at one and quickly
diverges. $p$ starts at one and remains almost constant. 
Typical black hole dyon solutions
are shown in Fig. \ref{BH_dyons}. 

\begin{figure}[bht]\centering
 \leavevmode 
\mbox{
\epsfxsize=8.cm \epsfbox{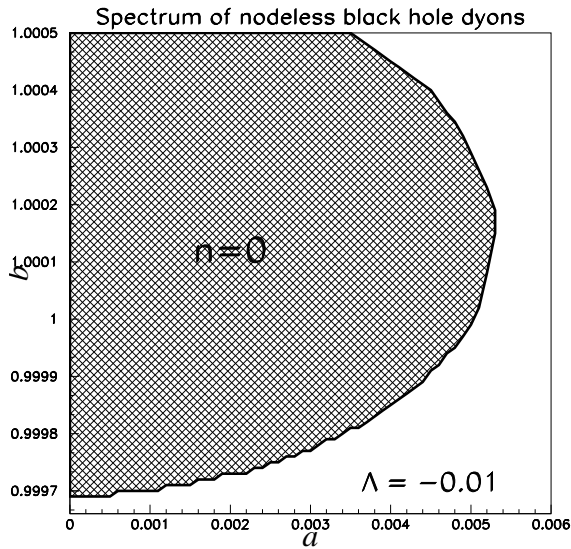}}
\caption{Spectrum of black hole solutions.   }
\label{BH_spectrum2}
\end{figure}

Again black hole dyon solutions with no node in $w(r)$ are stable
against small spherically symmetric perturbations.  The spectrum of
those nodeless black hole dyons in the parameter space $(a,b)$ is
plotted in fig.\ \ref{BH_spectrum2}.  Notice the   similarity
between fig.\ \ref{dyon-spectrum1} and fig.\ \ref{BH_spectrum2}.  The
nodeless solutions exist only for small $a=u_1 < 0.0055$.  $b=w_0$ must
be  around 1.

\section{Dependence on $\Lambda$ -- fractal structure}  

\begin{figure}[htb]
\centering \leavevmode 
\mbox{
\epsfxsize=8.cm \epsfbox{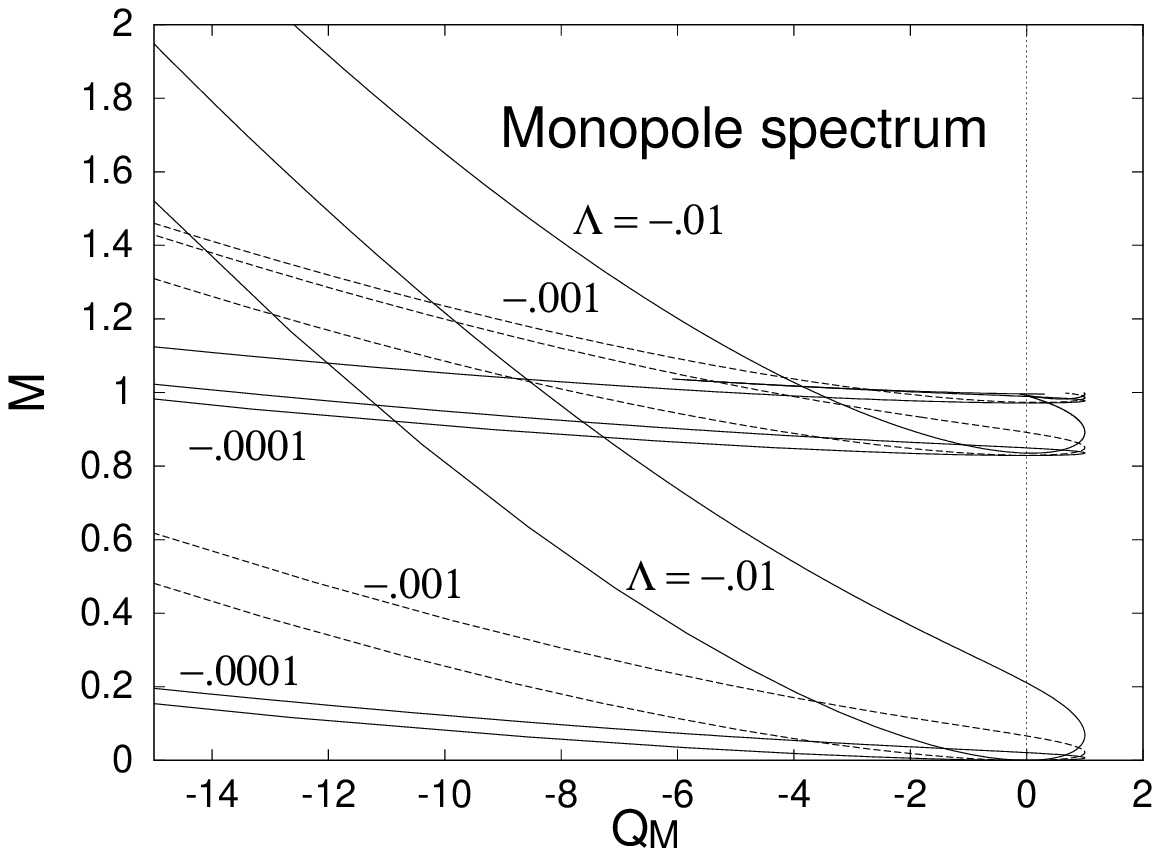}}
\vskip 0cm
\mbox{
\epsfxsize=8.cm \epsfbox{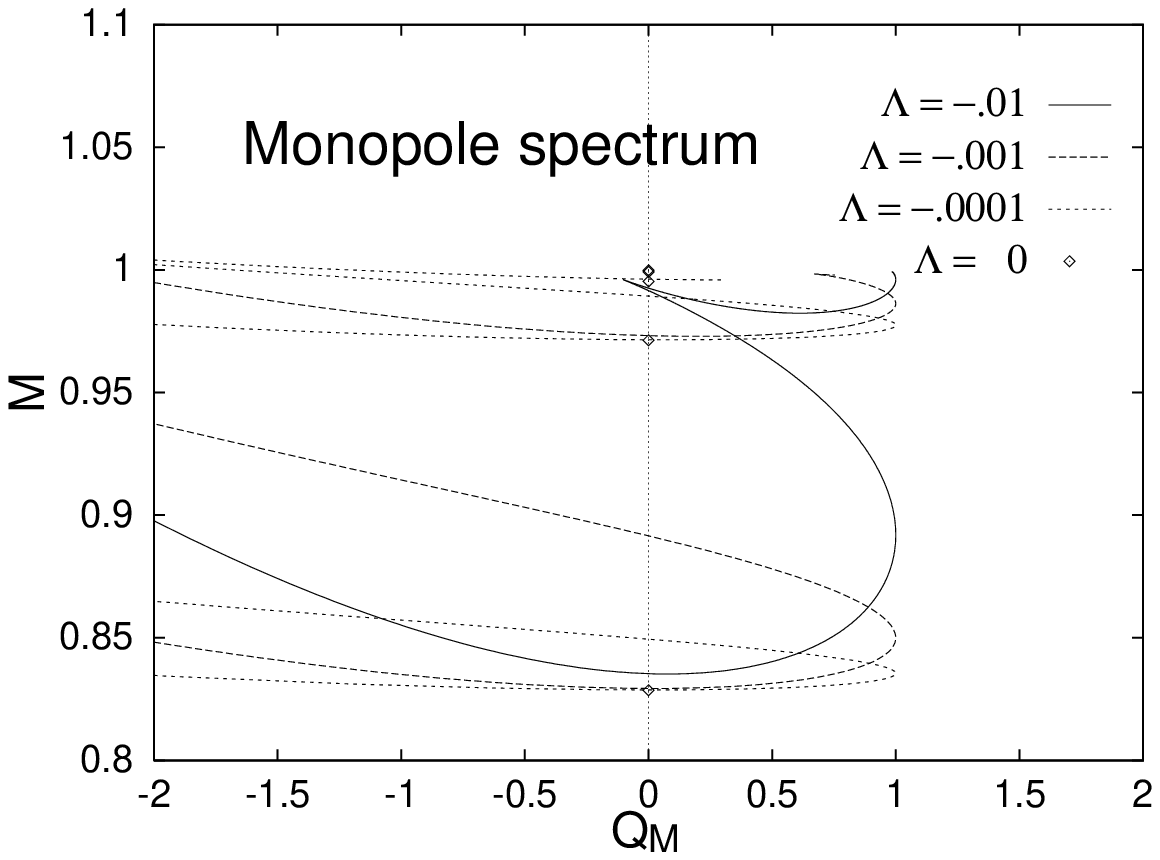}}
\caption{The appearance and collapse of soliton solutions in $\Lambda \go
0$.  The bottom figure is the blow-up of a part of the top figure
near $(Q_M, M)=(0,1)$, in which the first four Bartnik-McKinnon
solutions are also marked.  It is seen that as $\Lambda \go 0$, 
solutions on a branch collapse to a point.  New branches emerge as
$|\Lambda|$ becomes smaller, too.}
\label{monopole-Lamda}
\end{figure}

The soliton and black hole solutions depend  non-trivially
on the value of the cosmological constant $\Lambda$. It has not been well
understood why the continuum of solutions for negative $\Lambda$  become
a discrete set of solutions in the  $\Lambda\rightarrow 0$ limit, and
remain discrete for all
$\Lambda > 0$.  Just as fig.\ \ref{BH-spectrum1} shows for the black hole
solutions,  fig.\ 4 and fig.\ \ref{monopole-Lamda} shows the spectrum in
mass vs. magnetic charge $Q_M$ plane   for a give $\Lambda$.  The width
of each branch for a given $\Lambda$ gets smaller as $\Lambda$ approaches
zero. Fig.\ \ref{monopole-Lamda} indicates that as $\Lambda\rightarrow
0$, the branches collapse to one point, the BK solution, as the
continuum of solutions vanishes. It is still unknown mathematically why
and how this occurs.

\begin{figure}[htb]
\centering \leavevmode 
\mbox{
\epsfxsize=8.cm \epsfbox{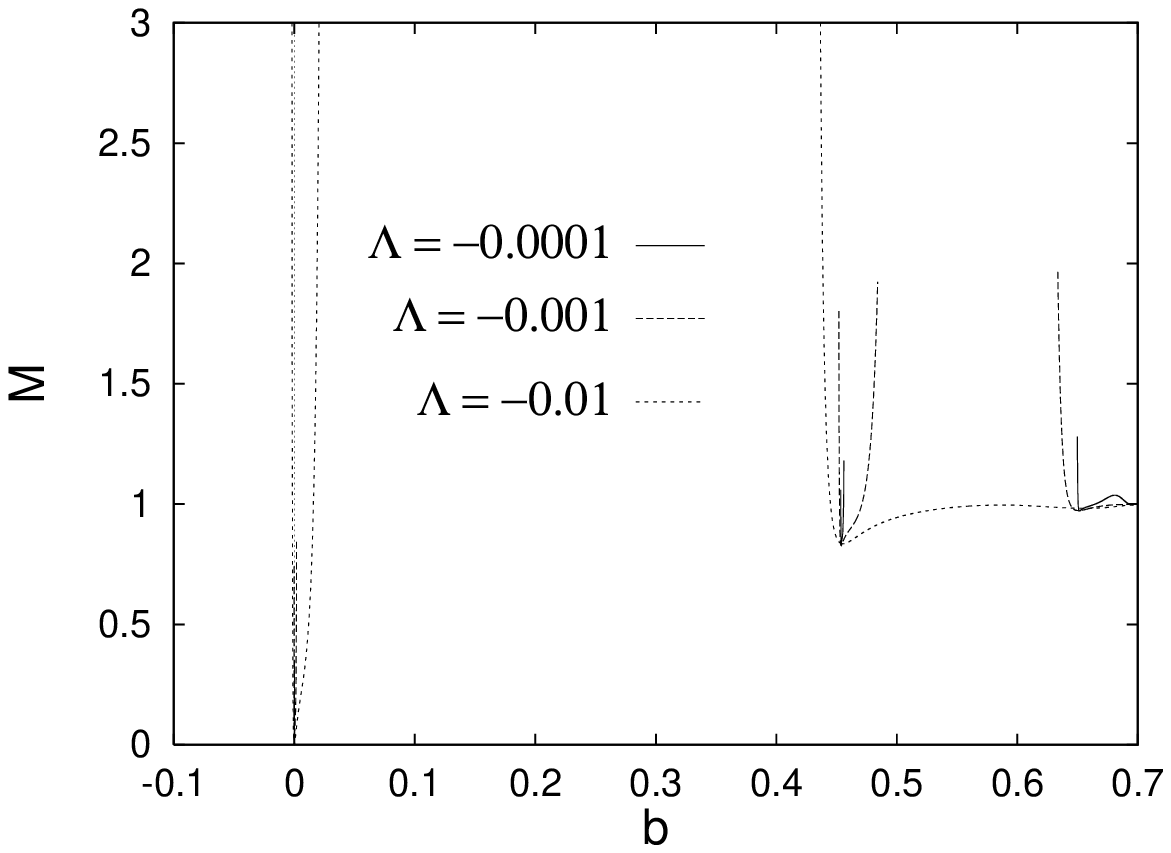}}
\caption{The monopole spectrum.  $b$ v.s. $M$ (mass) is plotted with
various values of $\Lambda$.}
\label{fractal1}
\end{figure}

\begin{figure}[htb]
\centering \leavevmode 
\mbox{
\epsfxsize=8.cm \epsfbox{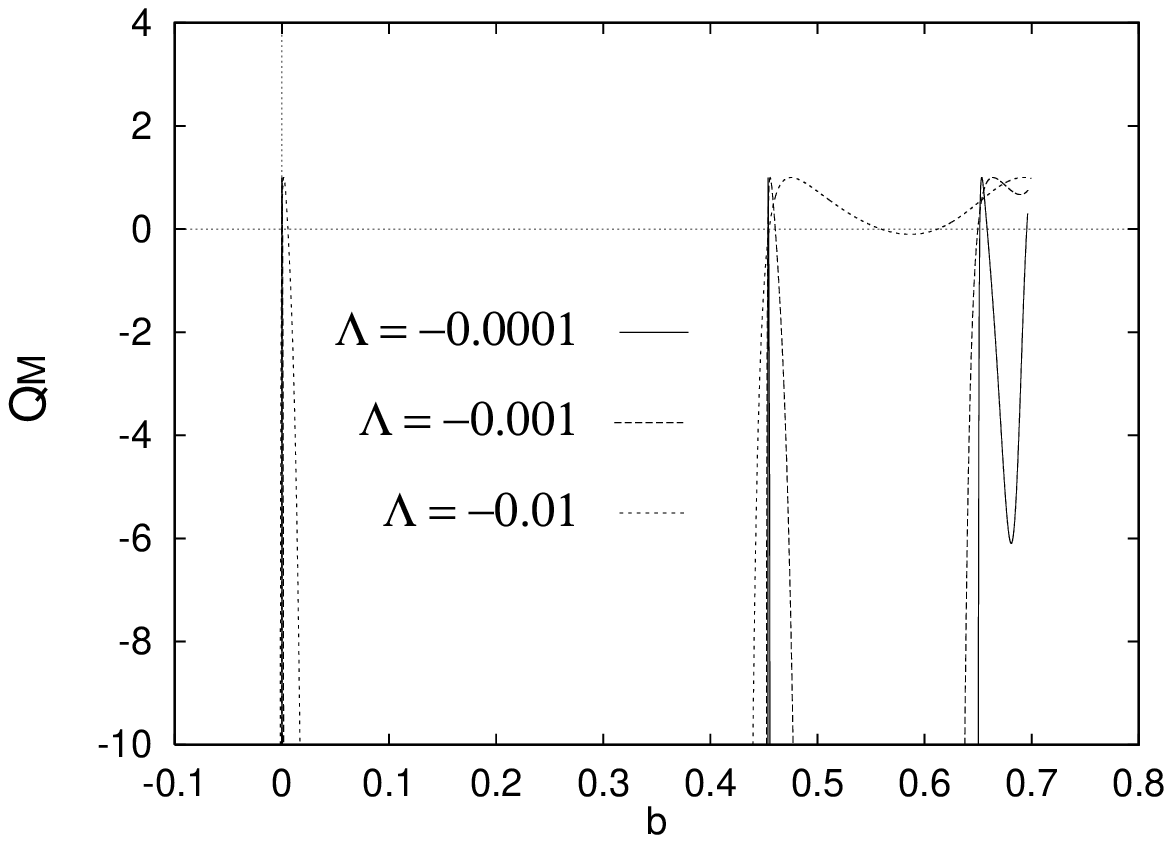}}
\caption{The monopole spectrum.  $b$ v.s. $Q_M$ (magnetic charge) is plotted
with various values of $\Lambda$.}
\label{fractal2}
\end{figure}

We would like to point out that there is a fractal structure in the 
moduli space of the solutions.   This is most clearly seen in the 
parameter $b$ v.s. mass $M$ plot as displayed in fig.\ \ref{fractal1}.
As $\Lambda$ becomes smaller, a new branch appears.  The shape 
of branches has approximate self-similarity.  Similarly,
in fig.\ \ref{fractal2} the magnetic charge $Q_M$ is plotted against $b$.
Delicate structure is observed near the critical $b=b_c$  which
signifies the critical solution discussed in Section 5.3 (c).
There may be some connection between the limiting point in the 
monopole spectrum and the critical solution.

\section{Stability}

It has been shown that the soliton and black hole solutions in
asymptotically Minkowski and de Sitter space, which necessarily have at
least one node in $w(r)$,  are unstable
\cite{ZHOU}\cite{GREENE}\cite{VOLKOV}\cite{VOLKOV2}. In contrast, 
the monopole and black hole solutions in the asymptotically  anti-de
Sitter space with no node in $w(r)$ are stable for
$u=0$. One expects the presence of the electric field not to change the
stability of the solitons and black hole configurations.  

In this section we give a detailed discussion for establishing the
stability.  We shall find that in   asymptotically anti-de Sitter
space the boundary condition for the resultant Schr\"odinger 
problem becomes subtle, and that the previous argument given in
asymptotically Minkowski space needs elaboration.

\subsection{Perturbation equations}

We consider small time-dependent perturbations to the static solutions  to
the coupled EYM equations. In the static solutions $\nu(r)=\tilde w(r)=0$.
In the general ansatz, (\ref{YM-ansatz3}) and (\ref{tetrads}) we set
\beqn
u(r,t) &=& u(r)+\delta u(r,t) \cr
w(r,t) &=& w(r)+\delta w(r,t) \cr
\tilde{w}(r,t) &=& \delta\tilde{w}(r,t) \cr
\nu(r,t) &=& \delta\nu(r,t) \cr
p(r,t) &=& p(r)+\delta p(r,t) \cr
H(r,t) &=& H(r)+\delta H(r,t)
\label{perturbation1}
\eeqn
and $m(r,t) = m(r) + \delta m(r,t)$.  Substituting (\ref{perturbation1}) 
into the Yang-Mills equation (\ref{YM-eq2}) and retaining only terms 
linear in perturbations, one finds
\beqn
&&\hskip -1.5cm
\Big[ r^2 u' \delta p +   r^2p (\delta u' - \delta \dot\nu) \Big]'
- {2uw^2p\over H} \bigg( {\delta p\over p} - {\delta H\over H} \bigg) 
-{2pw\over H} (w\delta u + 2 u\delta w + \delta \dot{\tilde w}) = 0 
       \label{YM_P_1} \\
\noalign{\kern 8pt}
&&\hskip -1.5cm
r^2p (\delta\dot{u}^{\prime} - \delta\ddot{\nu}) + r^2u' \delta\dot{p} -
\frac{2H}{p} \left[
w\delta\tilde{w}^{\prime}-\wpri\delta\tilde{w}+w^2\delta\nu
\right] =0 
      \label{YM_P_2} \\
\noalign{\kern 8pt}
&&\hskip -1.5cm
\Bigg[ {w'H\over p} \bigg( {\delta H\over H} - {\delta p\over p} \bigg) 
+ {H\over p} \delta w'  \Bigg]'
 +{pu\over H} (u \delta w + 2w\delta u + 2 \delta \dot{\tilde w})
  - {p\over H} \, \delta \ddot w   \cr
\noalign{\kern 8pt}
&&\hskip 2cm
+ {wu^2 p \over H} \bigg( {\delta p\over p} - {\delta  H\over H} \bigg) 
-\frac{w(1-w^2)}{r^2 p^2} \delta p
+ {1- 3w^2\over r^2 p} \delta w = 0  
\label{YM_P_3} \\
\noalign{\kern 8pt}
&&\hskip -1.5cm
\Bigg[ {H\over p} (\delta \tilde w' + w \delta \nu) \Bigg]'
+ {p\over H} (u^2 \delta \tilde w - 2u \delta \dot w -w\delta \dot u
   -  \delta \ddot{\tilde w} ) \cr
\noalign{\kern 8pt}
&&\hskip 2cm 
-{uwp\over H} \bigg( {\delta \dot p\over p} - {\delta \dot H\over H} \bigg) 
+ \frac{(1-w^2)}{r^2 p} \delta\tilde{w}
+\frac{H}{p} w' \delta\nu = 0~~ . 
\label{YM_P_4}
\eeqn
The Einstein  equations (\ref{Einstein2}) and (\ref{Einstein3}) yield
\beqn
&&\hskip -1.5cm
\delta m' = v\Bigg\{
r^2p^2\upri(\delta u' -\delta\dot{\nu})
+ \Big( r^2 u'^2 + {2u^2 w^2\over H} \Big) p \delta p 
+ \Big( w'^2 -\frac{p^2u^2w^2}{H^2} \Big) \delta H \cr
&&\hskip 1.5cm
 +2H w' \delta w'
+\frac{2uwp^2}{H}(w\delta u+u\delta w+\delta\dot{\tilde{w}}) 
-\frac{2w(1-w^2)}{r^2} \delta w  \Bigg\}
\label{EINSTEIN_M_1} \\
\noalign{\kern 8pt}
&&\hskip -1.5cm
\Bigg( {\delta p\over p} \Bigg)' = 
-{4v\over r} \Bigg\{
w' \delta w' + {p^2 u^2 w^2\over H^2} \bigg( {\delta p \over p} 
  - {\delta H\over H} \bigg) 
+ {p^2 uw\over H^2} (w\delta u + u \delta w + \delta \dot{\tilde w} )
   \Bigg\}
\label{EINSTEIN_P_1} \\  
\noalign{\kern 8pt}
&&\hskip -1.5cm
\delta \dot H = -{4vH\over r} 
\Big\{ w' \delta \dot w - u(w' \delta \tilde w - w \delta \tilde w')
 + uw^2 \delta \nu \Big\} ~~.  
\label{EINSTEIN_H_1}
\eeqn
There is residual gauge invariance specified by a gauge function 
$\Omega(r,t)$ in (\ref{gauge1}) and (\ref{YM-ansatz4}).  Making use of
this freedom, one can always set either $\delta u(r,t)=0$ or
$\delta \nu(r,t)=0$.

\subsection{Stability analysis}
In examining time-dependent fluctuations around monopole solutions
for which $u(r)=0$,  it is convenient to work in the $\delta u(t,r)=0$
gauge.   Eqs.\ (\ref{YM_P_1}), (\ref{YM_P_2}), and (\ref{YM_P_4})   
become
\beqn
&&\hskip -1cm
(r^2 p \delta \nu)' = - {2pw\over H} \, \delta \tilde w 
\label{YM_P_1a} \\
&&\hskip -1cm
r^2 p \delta \ddot \nu + {2H\over p} (w \delta \tilde w'
  - w' \delta \tilde w + w^2 \delta \nu) = 0
\label{YM_P_2a} \\
&&\hskip -1cm
\bigg[ {H\over p} (\delta \tilde w' + w \delta\nu) \bigg]'
 - {p\over H} \, \delta \ddot{\tilde w} 
 + {1-w^2\over r^2 p} \, \delta \tilde w 
 + {H\over p} \, w' \delta \nu = 0 ~~,
\label{YM_P_4a}  
\eeqn
whereas Eqs.\ (\ref{YM_P_3}), (\ref{EINSTEIN_M_1}), (\ref{EINSTEIN_P_1}),
and (\ref{EINSTEIN_H_1}) become
\beqn
&&\hskip -1cm
\Bigg[ {H\over p} \Bigg\{ w' 
\bigg( {\delta H\over H} - {\delta p \over p} \bigg)
   + \delta w' \Bigg\} \Bigg]' - {p\over H} \, \delta \ddot w
 - {w(1-w^2)\over r^2 p^2} \, \delta p 
 + {1-3w^2\over r^2 p} \, \delta w = 0
\label{YM_P_3a} \\
&&\hskip -1cm
\delta m' = v \Bigg\{  w'^2 \delta H + 2H w' \delta w'
 - {2w(1-w^2)\over r^2} \, \delta w \Bigg\}
\label{EINSTEIN_M_a} \\
&&\hskip -1cm
\bigg( {\delta p\over p} \bigg)' = - {4v\over r} \, w' \delta w' 
\label{EINSTEIN_P_a} \\
&&\hskip -1cm
\delta H = - {4v\over r} \, H w' \delta w ~~.
\label{EINSTEIN_H_a}
\eeqn
Notice that Eqs.\ (\ref{YM_P_1a}) - (\ref{YM_P_4a}) involve only
$\delta\nu$ and $\delta \tilde w$, defining the odd parity group,
whereas Eqs.\ (\ref{YM_P_3a}) - (\ref{EINSTEIN_H_a}) involve 
only $\delta w$, $\delta H$, and $\delta p$, defining the even parity
group.  The number of the equations is larger than the number of the 
unknown functions. Indeed, one equation in each group follows
from the others.

To derive the equation for each unknown function in a closed form,
we introduce the tortoise radial coordinate $\rho$ by
\beeq
{d\rho\over dr} = {p\over H}
\label{tortoise1}
\eneq
with which the equations for $w$, $p$, and $m$ become
\beqn
{d^2 w \over d\rho^2} &=& - {H\over r^2 p^2} \, w(1-w^2) 
   \label{tortoise_w} \\
\noalign{\kern 8pt}
{dp\over d\rho} ~ &=& - {2vp^2\over rH} \bigg( {dw\over d\rho} \bigg)^2
   \label{tortoise_p} \\
\noalign{\kern 8pt}
{dm\over d\rho} &=& v \Bigg\{
p \bigg( {dw\over d\rho} \bigg)^2 + {H\over 2r^2 p} \, (1-w^2)^2 
\Bigg\}  ~~.
\label{tortoise_m}
\eeqn
The range of $\rho$ is finite, $0 \le \rho \le \rho_{\rm max}$, 
since $p \go p_0$ and $H \go |\Lambda| r^2/3$ as $r\go\infty$:
\beeq
\rho = \cases{
r &for $r \sim 0$\cr
\noalign{\kern 8pt}
 \rho_{\rm max} - \myfrac{3p_0}{|\Lambda| r}  &for $r\sim \infty$ \cr}
\label{tortoise2}
\eneq
where $p=p_0 + {\rm O}(1/r)$.

In the odd parity group Eq.\ (\ref{YM_P_1a}) expresses $\delta w$
in terms of $\delta \nu$.  Substituting it into Eq.\ (\ref{YM_P_2a})
and making use of (\ref{tortoise_w}), one finds
\beqn
&&\hskip -1cm
\Bigg\{ - {d^2\over d\rho^2} + U_\beta(\rho) \Bigg\}  \beta
= \omega^2 \beta \cr
\noalign{\kern 8pt}
&&\hskip -1cm
U_\beta = \frac{H}{r^2p^2}(1+w^2)
+\frac{2}{w^2}\left(\frac{dw}{d\rho}\right)^2 ~~, \cr
\noalign{\kern 8pt}
&&\hskip -1cm
\delta\nu = {w\over r^2 p} \, \beta ~~~,~~~
\delta \tilde w = - {1\over 2w} \, {d\over d\rho} (w\beta) ~~.
\label{U_nu}
\eeqn
Here we have supposed fluctuations to be harmonic:
$\delta\nu(r,t)= e^{-i\omega t} \delta\nu(\rho)$ and 
$\delta\tilde w(r,t)= e^{-i\omega t} \delta\tilde w(\rho)$.
Eq.\ (\ref{YM_P_4a}) follows from (\ref{YM_P_1a}), (\ref{YM_P_2a}),
and (\ref{tortoise_w}).

In the even parity group, (\ref{EINSTEIN_P_a}) and (\ref{EINSTEIN_H_a})
express $\delta p$ and $\delta H$ ($\delta m$) in terms of $\delta w$.
Eq.\ (\ref{EINSTEIN_M_a}) automatically follows from (\ref{EINSTEIN_P_a}),
(\ref{EINSTEIN_H_a}), (\ref{tortoise_w}) and (\ref{tortoise_p}).
Eq.\ (\ref{YM_P_3a}) becomes, with the use of (\ref{EINSTEIN_P_a}),
(\ref{EINSTEIN_H_a}) and (\ref{tortoise_w}), 
\beqn
&&\hskip -1cm
\Bigg\{ - {d^2\over d\rho^2} + U_w(\rho) \Bigg\}  \delta w
= \omega^2 \delta w ~~, \cr
\noalign{\kern 8pt}
&&\hskip -1cm
U_w = \frac{H }{r^2 p^2} \, (3w^2-1)
+ 4 v \frac{d}{d\rho} 
  \bigg[ \frac{p}{rH} \Bigg( {dw\over d\rho} \bigg)^2 \Bigg]
\label{U_w}
\eeqn
Again harmonic fluctuations $\delta w(r,t)=e^{-i\omega t} \delta w(\rho)$
are supposed.

Eqs.\ (\ref{U_nu})  and (\ref{U_w}) have the same form as the
Schr\"odinger equation on a one-dimensional interval.  
Both of the potentials $U_\beta$ and $U_w$ are singular at
$\rho=0$,   behaving as $+ 2/\rho^2$.  $U_\beta$ has an additional
singularity if $w$ has  a zero at $\rho_k$; 
$U_\beta \sim + 2/(\rho -\rho_k)^2$.

The integrated energy-momentum density  $T^{ab} \sqrt{-g} \, d^3 x$ 
due to fluctuations must remain finite. At the origin $r=0$ it implies
that $\delta w=\delta \tilde w=0$ whereas $\delta \nu = {\rm O}(1)$. 
Taking advantage of the general coordinate invariance, one can impose
$\delta p=0$ and 
$\delta H=-2\delta m/r=0$.  At $r\sim \infty$, 
$\delta w', \delta \tilde w', \delta \nu = {\rm O}(r^{-2})$.
These are mild boundary conditions.  One can impose more strict
conditions such as the regularity at $r=0$ and vanishing 
at $r=\infty$.  As physical perturbations we demand that all $\delta w$,
$\delta \tilde w$, $\delta \nu$, $\delta H$, and $\delta p$ vanish
at $r=\infty$.

In Eq.\ (\ref{U_nu}) the potential $U_\beta(\rho)$ is positive definite.
However, this does not necessarily mean that the eigenvalue $\omega^2$
is positive definite.  It depends on the boundary condition.  Clearly
$\beta(0)=0$.  At $\rho=\rho_{\rm max}$, $\delta \nu=\delta \tilde w=0$
so that $\beta' + h \beta=0$ where $h=w'/w$.   Note that 
\beeq
h = {1 \over w}{dw\over d\rho} \Bigg|_{\rho_{\rm max}}
= {|\Lambda|\over 3p_0} \, {r^2\over w} {dw\over dr}\Bigg|_{r=\infty}
= {|\Lambda| w_1 \over 3p_0 w_0}
\label{boundary_h}
\eneq
where $w_j$'s are defined in (\ref{INFTY}).  For the monopole
configurations with no nodes in $w$, $h < 0$ ($h >0$) when $w$ is 
monotonically decreasing (increasing). 

Following Courant and Hilbert \cite{Courant}, we define
\beqn
&&\hskip -1.cm
{\cal D}(\vphi;h) =
 \int_0^{\rho_{\rm max}} d\rho \, \Big\{
 \vphi'(\rho)^2 +   U_\beta(\rho) \, \vphi(\rho)^2 \Big\}
+ h ~ \vphi(\rho_{\rm max})^2 \cr
\noalign{\kern 10pt}
&&\hskip -1.cm
{\cal N}(\vphi) = 
  \int_0^{\rho_{\rm max}} d\rho \, \vphi(\rho)^2  ~~.
\label{variation1}
\eeqn
If $w(r)$ is nodeless, then $U_\beta(\rho)$ is regular on the 
interval except at $\rho=0$.  The equation implies that
$\beta={\rm O}(\rho^2)$ near the origin. In this case, for an
eigenfunction $\beta(\rho)$ in (\ref{U_nu}) satisfies
\beeq
\omega^2 = {{\cal D}(\beta) \over {\cal N}(\beta)} ~~.
\label{variation2}
\eneq
It follows immediately that all eigenvalues $\omega^2$ are positive
definite if $h \ge 0$ so that the solution is stable against small
odd-parity perturbations.

For $h<0$ more careful analysis is necessary.  
The lowest eigenvalue
$\omega^2 \equiv \lambda_1$ in the eigenvalue equation (\ref{U_nu}) is 
exactly the lower bound of the set of values assumed by the functional
${\cal D}(\vphi,h)$, where $\vphi$ is any function continuous 
on the interval $[0,\rho_{\rm max}]$ with piecewise continuous derivatives
satisfying $\vphi(0)=1$ and ${\cal N}(\vphi)=1$.  
\beeq
\lambda_1(h) = {\rm min}_\vphi ~ 
   {\cal D}(\vphi;h) ~~.
\label{variation3}
\eneq
If $\lambda_1 >0$, then the solution is stable against odd-parity
perturbations.
Suppose that $\vphi_1(\rho)$ saturates the lower bound for 
$h_1$: $\lambda_1(h_1) = {\cal D}(\vphi_1;h_1)$.  As
\beqn
\lambda_1(h_1) &=& {\cal D}(\vphi_1;h_2) 
   + (h_1 - h_2) \vphi_1(\rho_{\rm max})^2 \cr
\noalign{\kern 8pt}
&\ge& \lambda_1(h_2) +  (h_1 - h_2) \vphi_1(\rho_{\rm max})^2 ~,
\label{variation4}
\eeqn
$\lambda_1(h)$ is a monotonically increasing function of $h$. 
Hence, if $\lambda_1(h_1)>0$, then  $\lambda_1(h) > 0$ for $h \ge h_1$.

To establish the stability we utilize the residual
gauge invariance.  
There is a zero-mode (with $\omega^2=0$) for Eq.\  (\ref{U_nu})
with an appropriate boundary condition $h_0$.
In the $\Lambda=0$ case the existence of the zero-mode was
utilized to prove the instability of the BK and black hole solutions which
has at least one node in $w(r)$ \cite{VOLKOV3}\cite{Kanti}.  Consider
the time-independent gauge function $\Omega(r)$ in (\ref{gauge1}).
For $|\Omega| \ll 1$, $\delta\nu= d\Omega/dr$ and 
$\delta\tilde w = - w\Omega$.    Eq.\ (\ref{YM_P_1a}) is satisfied
if
\beeq
{d\over d\rho} \bigg({r^2 p^2 \over H} {d \Omega\over d\rho} \bigg)
=  2 w^2 \Omega ~~.
\label{zero-mode1}
\eneq
As $\Omega(0)=0$, $\Omega \sim a \rho + {\rm O}(\rho^3)$ 
for $\rho \sim 0$.  Hence Eq.\ (\ref{zero-mode1}) determines 
$\Omega(r)$ up to an over-all constant.  
$\beta_0= (r^2 p^2 / wH) (d\Omega/d\rho)$ is the zero mode
of Eq.\ (\ref{U_nu}) and $U_\beta=\beta_0''/\beta_0$,    where $\Omega'
\equiv d\Omega/d\rho$ etc.  In this case $\Omega(\rho_{\rm max}) \not= 0$
and $h_0 = - \beta_0'/\beta_0 |_{\rho_{\rm max}}$ differs from
$h$ in the eigenvalue problem under consideration.  If $h > h_0$,
then $\lambda_1(h) > 0$, establishing the stability.  
As $d(w\beta_0)/d\rho=2w^2 \Omega$, 
\beeq
h = h_0 + {2w^2 H \Omega\over r^2 p^2 \Omega'} ~~.
\label{boundary_h1}
\eneq
Nodeless solutions ($w>0$) are stable if $\Omega/\Omega' >0$
at $\rho_{\rm max}$.

Solving (\ref{zero-mode1}) numerically,  we have determined
$\Omega(\rho)$ to find that indeed $h > h_0$ for nodeless
solutions.  This analysis also shows that $h$ becomes exactly $h_0$ for
the  configuration with $w(r=\infty)=0$. In this limiting case
the zero mode is not normalizable; it diverges as 
$(\rho-\rho_{\rm max})^{-1}$.

This is a general behavior.  When $w$ has a node at  
$\rho_k < \rho_{\rm max}$,  there appears a negative $\omega^2$ mode which
behaves as $(\rho - \rho_k)^{-1}$ near $\rho_k$.

If $w(r)$ has $n$ nodes, i.e.
$w(r_j)=0$ ($j=1, \cdots, n$), the potential $U_\beta$ develops
$(\rho-\rho_j)^{-2}$ singularities.   Volkov et al.\ have shown for the 
BK solutions in the $\Lambda=0$ case that there appear exactly $n$
negative eigenmodes ($\omega^2<0$) if $w$ has $n$ nodes \cite{VOLKOV3}. 
A similar conclusion has been obtained for black hole solutions as well
\cite{Kanti}. Their argument needs elaboration in the $\Lambda <0$ case,
however.  

To investigate the eigenvalue spectrum of (\ref{U_nu}) in this case, it is
convenient to consider the dual equation as was done in \cite{VOLKOV3}. 
One can write the Schr\"odinger equation in (\ref{U_nu}) as
\beqn
&&\hskip -1.cm
Q_+ Q_- \vphi_n = \lambda_n \vphi_n \cr
\noalign{\kern 10pt}
&&\hskip -1.cm
Q_\pm = \pm {d\over d\rho} + {\beta_0'\over \beta_0} 
\label{dual1}
\eeqn
where $\beta_0(\rho)$ is the zero mode described above.  
$\vphi_n={\rm O}(\rho^2)$ near $\rho=0$ and 
$\vphi_n' + h \vphi_n=0$ at $\rho_{\rm max}$.  The dual
equation is given by
\beqn
&&\hskip -1.cm
Q_- Q_+ \tilde\vphi_n 
= \bigg\{  - {d^2\over d\rho^2} + \tilde U_\beta \bigg\} \tilde\vphi_n 
= \lambda_n \tilde\vphi_n \cr
\noalign{\kern 10pt}
&&\hskip -1.cm
\tilde U_\beta =  - {H\over r^2 p^2} (1+w^2) 
 - {8w w' H\Omega\over r^2 p^2 \Omega'}
 + 8 \bigg( {w^2 H \Omega\over r^2 p^2 \Omega'} \bigg)^2 ~~.
\label{dual2}
\eeqn
$\vphi_n$ and $\tilde\vphi_n$ are related to each other by
\beeq
\tilde\vphi_n = \cases{
 Q_- \vphi_n  &for $\lambda_n \not= 0$\cr
{\vphi_n}^{-1}=  {\beta_0}^{-1} &for $\lambda_n = 0$ \cr}
\label{dual3}
\eneq
However, the boundary condition for $\tilde\vphi_n(\rho)$ depends on
$\lambda_n$:
\beqn
&&\hskip -1.cm
\tilde\vphi_n(0) = 0 \cr
\noalign{\kern 5pt}
&&\hskip -1.cm
\tilde\vphi_n'(\rho_{\rm max}) 
  + \tilde h_n \tilde\vphi_n(\rho_{\rm max}) =0 \cr
\noalign{\kern 5pt}
&&\hskip -1.cm
\tilde h_n  = - h_0 - {\lambda_n \over h - h_0}  ~~.
\label{dual4}
\eeqn

\begin{figure}[htb]
\centering \leavevmode 
\mbox{
\epsfxsize=8.cm \epsfbox{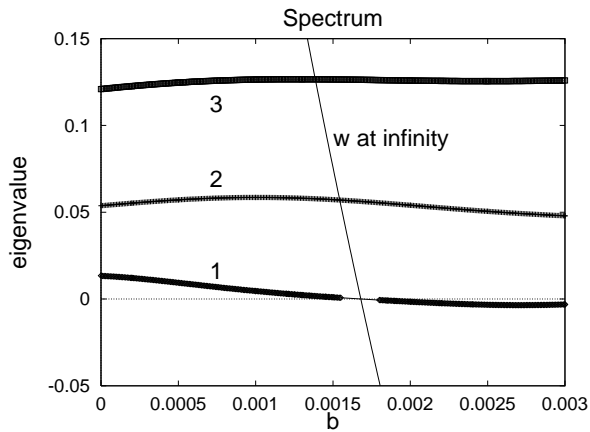}}
\caption{Eigenvalues $\lambda_n$ ($n=1,2,3$) in the dual Schr\"odinger
equation (\ref{dual2}) with the boundary condition (\ref{dual4})
are displayed for monopole configurations with varying value of $b$.  
$\Lambda=-0.01$ and $v=1$. When
$w(\infty)$ develops a node, the negative eigenvalue mode arises,
signaling the instability of the solution.  
$w(\infty)=0$ at $b=0.00168$,  where the numerical evaluation of
$\lambda_1$ becomes  difficult.}
\end{figure}

The advantage of considering the dual equation is that the dual
potential $\tilde U_\beta(\rho)$ is regular except at $\rho=0$
where it behaves as $+ 6/\rho^2$.  However, the eigenvalue $\lambda_n$
has to be determined self-consistently such that the boundary
condition (\ref{dual4}) is satisfied.  We have determined $\lambda_n$'s
numerically for the monopole configurations in the lower branch in fig.\ 4.
The first, second and third eigenvalues are displayed in fig.\ 14.
One sees that $\lambda_1 >0$ for the nodeless configurations, but
the unstable mode develops when $w$ has a node.  

The wave function 
of the unstable mode in the original equation, not in the dual
equation, diverges at the zeroes of $w(r)$.  In other words,
the instability sets in around the zeroes of $w(r)$.
The potential $U_\beta(\rho)$ and $\tilde U_\beta(\rho)$
for the solution at $b=0.0025$  are plotted in fig.\ 15.
At the node of $w$, $U_\beta$ diverges, but $\tilde U_\beta$ remains
finite.  The wave function $\tilde\vphi_1(\rho)$ of the lowest eigenvalue 
($\lambda_1 = -0.0033)$ and the corresponding $\vphi_1(\rho)$ also
have been plotted in fig.\ 15.

\begin{figure}[htb]
\centering \leavevmode 
\mbox{
\epsfxsize=8.cm \epsfbox{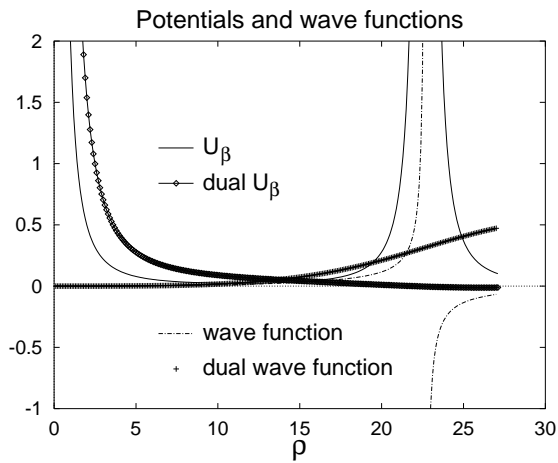}}
\caption{$U_\beta(\rho)$ and $\tilde U_\beta(\rho)$ for the monopole
solution at $b=0.0025$, $\Lambda=-0.01$, and $v=1$.  The wave function of
the unstable mode ($\lambda=-0.0033$) are also plotted. 
$\tilde\vphi_1(\rho)$ in the dual equation (\ref{dual2})
is a regular smooth function, but $\vphi_1(\rho)$
in the original equation  (\ref{dual1}) diverges at the zero of $w$.}
\end{figure}

For even-parity perturbation  the potential
$U_w(\rho)$  in Eq.\ (\ref{U_w}) is not positive definite.  The first term
in  $U_w$ becomes negative for
$w^2 < 1/3$.  The second term also  can become negative when $w'$
vanishes at finite
$r$.   We have solved the  Schr\"odinger equation (\ref{U_w}) numerically
for typical monopole solutions, and found that for the solutions with 
no node in $w(r)$, the eigenvalues $\omega^2$ are always positive
even if $w(r=\infty)< 1/\sqrt{3}$.

Hence we have established the stability of the monopole solutions
with no node in $w(r)$.

\section{Summary}

New monopole, dyon, and black hole  solutions to the Einstein-Yang-Mills
equations have been found in asymptotically anti-de Sitter space.
The solutions with no node in the non-Abelian field strengths are
shown to be stable against spherically symmetric perturbations.
The non-trivial boundary condition plays a crucial role in developing
the instability for solutions with nodes.  The stability of nodeless dyon
solutions need to be established.

Though electric and magnetic charges of monopole and dyon solutions
are not quantized in classical theory, they are expected to be quantized
in quantum theory.  If this is the case, then at least solutions
with the smallest charge would become absolutely stable.

We have also found the critical spacetime solutions which end at
finite $r$.  These solutions may have connections to black hole
solutions, though more detailed study is necessary.

The solutions  found in the present paper may have profound consequences
in the  evolution of the early universe which may have gone through
the anti-de Sitter phase.  We hope to report on these subjects
in future publications.

\vskip 1cm

\leftline{\bf Acknowledgments}

This work was supported in part by the U.S.\ Department of
Energy under contracts DE-FG02-94ER-40823, DE-FG02-87ER40328
and DE-AC02-98CH10886.

\end{document}